\documentclass[aps,prb,reprint,longbibliography,superscriptaddress]{revtex4-2}
\usepackage{mathrsfs}
\usepackage{amsmath,gensymb}
\usepackage{amsfonts}
\usepackage{amssymb}
\usepackage{amsthm}
\usepackage{graphicx}
\usepackage{natbib}
\usepackage{color}
\usepackage{hyperref}
\usepackage{bm}
\usepackage[caption=false]{subfig}
\usepackage{verbatim}
\usepackage[normalem]{ulem}
\usepackage{soul}

\begin{document}
\title{Spin supercurrent in superconductor/ferromagnet van-der-Waals heterostructures}

\author{G. A. Bobkov}
\affiliation{Moscow Institute of Physics and Technology, Dolgoprudny, 141700 Moscow region, Russia}

\author{A. M. Bobkov}
\affiliation{Moscow Institute of Physics and Technology, Dolgoprudny, 141700 Moscow region, Russia}

\author{I.V. Bobkova}
\affiliation{Moscow Institute of Physics and Technology, Dolgoprudny, 141700 Moscow region, Russia}
\affiliation{National Research University Higher School of Economics, 101000 Moscow, Russia}

\begin{abstract}
We study dissipationless spin transport induced by a charge supercurrent in a monolayer van der Waals superconductor  under the applied magnetic field and in a bilayer superconductor/ferromagnet (S/F) heterostructure with no external field. It is shown that in both cases a combined action of the Ising-type spin-orbit coupling  and the Zeeman field results in the appearance of nonunitary superconducting triplet correlations with nonzero average Cooper pair spin, which carry spin current in the presence of a condensate motion. Properties of this dissipationless spin current are investigated. In particular, it is shown that it manifests a rectification effect. In addition, in S/F heterostructures the amplitude and the sign of the spin current are controlled by gating.
\end{abstract}

\maketitle

\section{Introduction}

In spintronics, spin current plays an essential role to transfer
the information associated with the spin degrees of freedom and to manipulate the magnetization. Spin currents can be carried by itinerant electrons or spin waves. In the first case they can be generated via various methods including the spin-polarized current
injection from the ferromagnet\cite{Datta1990,Gardelis99,Schmidt00}, spin battery\cite{Saitoh2006,Wei2014,Dushenko2016,Lesne2016,Kondou2016}, spin Hall
effect\cite{Hirsch99,Zhang00,Murakami2003,Sinova04,Kato2004,Wunderlich05,Valenzuela2006}, by applying an electric field to non-centrosymmetric systems with spin-orbit coupling\cite{Yu2014,Hamamoto2017}. The spin wave spin currents are carried by excitations of localized magnetic moments through exchange interactions and can be implemented even in insulating ferromagnets \cite{Barman2021,Democritov2013,Rezende2020,Brataas2020,Chumak2015}. 

The spin currents mentioned above are accompanied by energy dissipation either through the Joule heat or through damping in the magnetic system. On the other hand, it would be desirable to be able to generate and transport non-dissipative or weakly dissipative spin currents over long distances. This opportunity is provided by superconducting spintronics \cite{Linder2015,Eschrig2015}. One of the directions of the superconducting spintronics is the study of possible ways to generate and control spin currents, the carriers of which are Cooper pairs. Although the spin degrees of
freedom usually do not manifest themselves in singlet superconductors, the superconducting spin current carried by triplet Cooper pairs can be finite in $\mathrm{^3He}$ \cite{Leggett1975}, triplet superconductors \cite{Asano2005,Asano2006}, noncentrosymmetric superconductors \cite{Leurs2008,He2019,Bergeret2016}, as well as in superconductor/ferromagnet (S/F) hybrids and Josephson junctions via ferromagnetic weak links \cite{Grein2009,Jeon2020,Brydon2013,Dahir2022,Ojajarvi2022,Ouassou2017,Ouassou2019,Alidoust2010,Jacobsen2016,Gomperud2015,Brydon2011,Linder2017,Bobkova2017,Bobkova2018,Aunsmo2024}. In particular, it was shown that in two-dimensional superconductors with Rashba spin-orbit interaction, the
generation of dissipationless bulk spin current by charge supercurrent is possible \cite{He2019}. The charge supercurrent induces an averaged spin polarization of the pairs via the Edelstein effect \cite{Edelstein1995,Edelstein2005,Ilic2020}, which is linear in the supercurrent and the spin current is just the charge current times the spin polarization. Consequently, spin supercurrent is proportional to the charge supercurrent squared. A very large number of important works are devoted to spin currents in S/F heterostructures \cite{Grein2009,Jeon2020,Brydon2013,Dahir2022,Ojajarvi2022,Ouassou2017,Ouassou2019,Alidoust2010,Jacobsen2016,Gomperud2015,Brydon2011,Linder2017,Bobkova2017,Bobkova2018,Aunsmo2024}. However, they mostly study spin supercurrents, which exist only in restricted regions of space, for example weak links of S/F/S Josephson junctions or regions of the order of the superconducting coherence length near the interfaces of S/F and F/S/F structures. The main interest of such spin supercurrents is their capability to induce spin transfer torques and magnetization dynamics of the ferromagnetic elements \cite{Zhu2004,Zhu2005,Holmqvist2011,Linder2011,Halterman2016,Kulagina2014,Linder2012,Takashima2017}.   

From the other hand, quantum materials with a large controllable charge current-induced spin polarization and capability to transfer spin currents over long distances are promising for next-generation all-electrical spintronic science and technology \cite{Shrivastava2021,Geim2013}. Van der Waals (vdW) metals with high spin-orbit coupling have attracted significant attention for an efficient charge to spin conversion process \cite{Khokhriakov2020,Lin2019,Guimaraes2018,Shi2019,MacNeill2017,
Zhao2020,Stiehl2019,Safeer2019,Zhao2020_WTe2,Ghiasi2019}. In particular, an electrical generation of spin polarization in $\mathrm{NbSe_2}$ up to room temperature has been demonstrated \cite{Hoque2022}. However, in that study the charge-spin conversion signal was only observed with a higher bias current above the superconducting critical current, limiting the observation of signal only to the non-superconducting state of $\mathrm{NbSe_2}$. 

Here we predict that charge current-induced spin supercurrents can occur in superconducting $\mathrm{NbSe_2}$ monolayers under the applied magnetic field or in bilayer $\mathrm{NbSe_2}$/F heterostructures, where F is a  vdW ferromagnet consisting of one or a few layers. We  unveil the physical role of the Ising-type spin-orbit coupling and the Zeeman field in the generation of non-unitary triplet Cooper pairs possessing  nonzero averaged spin, which become spin current carriers in the presence of the condensate motion. It is also obtained that the pair spin current has a component, which manifests a rectification effect unlike the charge supercurrent that generates it. For the case of S/F bilayers it is shown that the degree and sign of the spin polarization can be controlled by gating.  

The paper is organized as follows. In Sec.~\ref{sec:applied_field} we present analytical and numerical results for the spin currrent in $\mathrm{NbSe_2}$ monolayer under the applied external magnetic field. Sec.~\ref{sec:field_model} is devoted to formulation of the model, and in Sec.~\ref{sec:field_method} we describe the Green's function approach, which is used further for analytical calculations. In Sec.~\ref{sec:field_spin} we present analytical results for the averaged spin of non-unitary Cooper pairs, which arise as a result of partial conversion of singlet pairs under the influence of the Ising spin-orbit coupling and the applied field. Analytical results for the spin current carried by these pairs are presented in Sec.~\ref{sec:field_current_analytical}. Sec.~\ref{sec:field_current_numerical} is devoted to the numerical results for the spin current, which go beyond our analytical approximation and exhibit important additional physics including rectification effect.  In Sec.~\ref{heterostructure} we study spin current in a S/F heterostructure on a specific example of $\mathrm{NbSe_2/VSe_2}$ heterostructure. Sec.~\ref{sec:heterostructure_model} is devoted to the description of the corresponding model and in Sec.~\ref{sec:heterostructure_current} the results for the spin current in this heterostructure are presented, including demonstration of spin current control using gating potential. Sec.~\ref{conclusions} contains conclusions from our work. 

\section{$\mathrm {NbSe_2}$ under the applied magnetic field}

\label{sec:applied_field}

\subsection{Model}

\label{sec:field_model}

\begin{figure}[tb]
	\begin{center}
		\includegraphics[width=85mm]{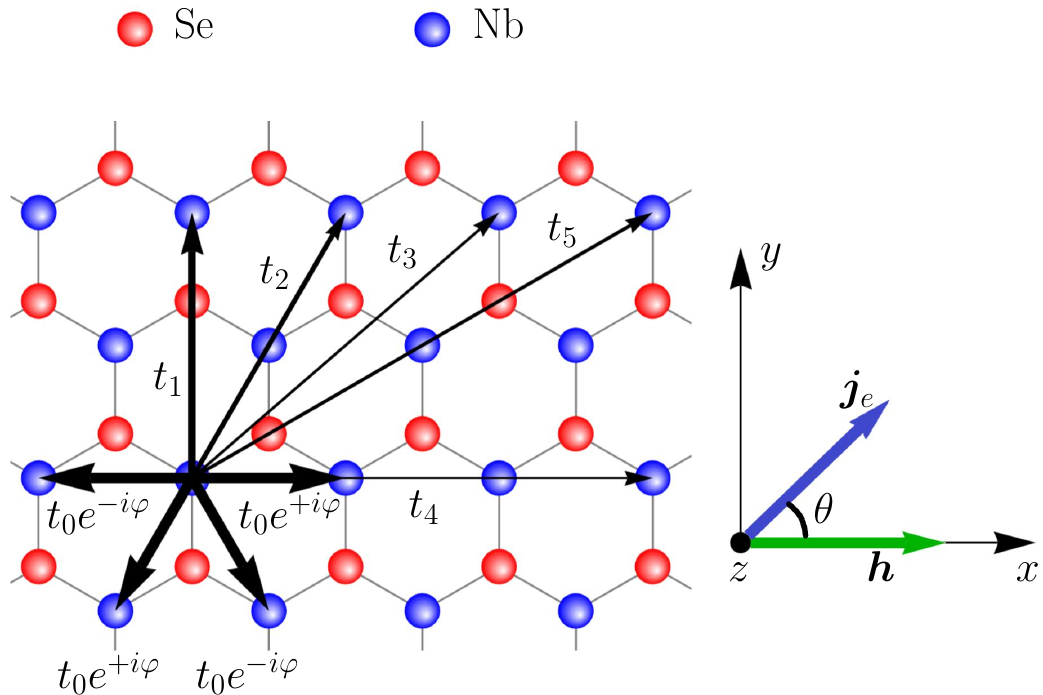}
\caption{Top view of the $\mathrm{NbSe_2}$ monolayer atomic structure. The electronic hops involved in the tight-binding Hamiltonian (\ref{eq:hamiltonian}) are shown by black arrows. Bold black arrows correspond to the nearest-neighbor hopping. The phases of these hopping elements accounting for the Ising-type spin-orbit coupling are also shown. It is assumed that a Zeeman field $\bm h$ and a supercurrent $\bm j_e$ are applied to the system.}
 \label{fig:sketch}
	\end{center}
 \end{figure}

\begin{figure}[tb]
	\begin{center}
		\includegraphics[width=85mm]{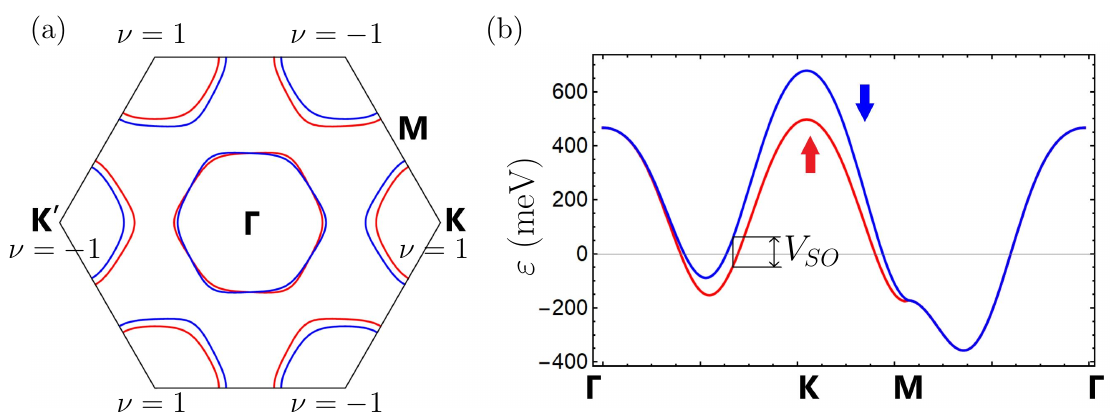}
\caption{(a) Fermi surface of the electronic system described by Hamiltonian (\ref{eq:hamiltonian}) with parameters listed in Tab.~\ref{tab:hopping_par}. The valley index  corresponding to each of the contours around $K$-points $\nu = \pm 1$ is indicated. Red contours are related  to spin $\uparrow$ ($s=+1$), and blue contours correspond to spin $\downarrow$ ($s=-1$). (b) DFT-calculated low-energy band structures of the 1H-$\mathrm{NbSe_2}$ monolayer. See Ref.~\onlinecite{Bobkov2024_vdW} for parameters and details of calculations.}
 \label{fig:fermi}
	\end{center}
 \end{figure}

We consider a $\mathrm {NbSe_2}$ monolayer under the applied magnetic field. The atomic structure of the monolayer is
illustrated in Fig.~\ref{fig:sketch}. The electronic structure of the system is modelled by the following tight-binding Hamiltonian:
\begin{eqnarray}
\hat H = - \sum\limits_{<\bm i\bm j>,\sigma} c^\dagger_{\bm i,\sigma}t_S^{\bm i\bm j,\sigma} c_{\bm j,\sigma}-
\mu_S\sum\limits_{\bm i,\sigma} c^\dagger_{\bm i,\sigma} c_{\bm i,\sigma}+ \nonumber \\
\sum\limits_{\bm i,\alpha,\beta} c^\dagger_{\bm i,\alpha}(\bm h\bm\sigma)_{\alpha,\beta} c_{\bm i,\beta}+
\sum\limits_{\bm i}\left[ \Delta c_{\bm i,\uparrow} c_{\bm i,\downarrow}+H.c.\right]
\label{eq:hamiltonian}
\end{eqnarray}
Here $c_{\bm i,\sigma} $ is an electron annihilation operator at site $\bm i$ in plane of the $\mathrm {NbSe_2}$ layer and for spin $\sigma = \uparrow, \downarrow$. $\mu_{S}$ is a chemical potential of the $\mathrm {NbSe_2}$ layer. To correctly describe electronic structure of the $\mathrm {NbSe_2}$ monolayer we assume not only nearest-neighbor hopping, but several hops $\bm i\to \bm j$ with complex hopping elements $t^{\bm i\bm j,\sigma}_S$. $\langle \bm i\bm j \rangle$ means summation over all involved neighbors. To find the values of these hopping elements the DFT-calculated low-energy electron spectra were fitted by a single-band tight-binding model in a triangular lattice \cite{Bobkov2024_vdW}, taking into account the complex hopping elements between the first to the sixth neighbours $t_S^0 e^{i\varphi_0}, ..., t_S^5 e^{i\varphi_5}$, where $t_S^i$ is the corresponding hopping energy and $\varphi_i$ accounts for the spin-orbit coupling. Only $\varphi_0 \equiv \varphi \neq 0$ for the case of the Ising-type spin-orbit coupling. The resulting spin splitting of the normal state electronic spectra of $\mathrm {NbSe_2}$ in the vicinity of the Fermi surface around $K$-point is denoted as $V_{SO}$. Its exact value depends on the particular momentum point, but approximately  $V_{SO} \approx 6 t_0 \sin \varphi$. It is rather large $V_{SO} \sim 100 $meV and is shown  in Fig.~\ref{fig:fermi}.  All the involved hops are schematically shown in Fig.~\ref{fig:sketch} by black arrows. The parameters extracted from the fits are listed in Table~\ref{tab:hopping_par}. They are in good agreement with the data reported earlier \cite{Aikebaier2022}. 

\begin{table}
\begin{center}
\begin{tabular}{|c|c|c|c|c|c|c|c|}
\hline
 $\mu_S$ & $t_S^0$ & $t_S^1$ & $t_S^2$ & $t_S^3$ & $t_S^4$ & $t_S^5$ & $\varphi$  \\
\hline
31.4 & 17.5 & 99.8 & -7.8 & -3.6 & -14.3 & 0.5 & 1.48  \\
\hline
\end{tabular}
\end{center}
\caption{\label{tab:hopping_par} Parameters of the one-band tight-binding model fitted to the DFT-calculated electron spectra of $\mathrm {NbSe_2}$. All values of the hopping amplitudes and other energies are given in meV.}
\end{table}

$\bm h$ is the Zeeman field produced by the applied magnetic field. The orbital effect of the applied magnetic field is small in the monolayer limit and can be neglected.  $\Delta$ is the superconducting order parameter, which is to be calculated self-consistently as $\Delta = \lambda \langle c_{\bm i,\downarrow} c_{\bm i,\uparrow} \rangle$, where $\lambda$ is the pairing constant. Then we apply a supercurrent $\bm j_e$ along the $\mathrm{NbSe_2}$ layer and study the spin supercurrent generated by this charge supercurrent.

\subsection{Green's functions technique}

\label{sec:field_method}

For analytical calculation of the amplitudes of triplet Cooper pair correlations, average spin of the Cooper pair and the spin current we use the Green's functions technique. The Matsubara Green's function is $4 \times 4$ matrix in the direct product of spin and particle-hole spaces. Introducing the Nambu spinor $\check \psi_{\bm i} = (c_{{\bm i},\uparrow}, c_{\bm i,\downarrow}, c_{\bm i,\uparrow}^{\dagger}, c_{\bm i,\downarrow}^{\dagger}, )^T$ we define the Green's function as follows: 

\begin{eqnarray}
\check G_{\bm i \bm j}(\tau_1, \tau_2) = - \langle T_\tau \check \psi_{\bm i}(\tau_1) \check \psi_{\bm j}^\dagger(\tau_2) \rangle 
\label{Green_Gorkov}
\end{eqnarray}
where $\langle T_\tau ... \rangle$ means  imaginary time-ordered thermal averaging. Introducing Pauli matrices in spin, and particle-hole spaces: $\sigma_k$ and $\tau_k$  ($k=0,x,y,z$) and operator $\hat j $ as
\begin{eqnarray}
\hat j c_{\bm i,\sigma} = \sum\limits_{< \bm i \bm j>}t_S^{ij,\sigma} c_{\bm j,\sigma}
\label{op_j}
\end{eqnarray}
one can obtain the Gor'kov equation for the Green's function in terms of the Matsubara frequencies $\omega_m = \pi T(2m+1)$. The derivation is similar to that described in Ref.~\onlinecite{Bobkov2022}. The resulting Gor'kov equation takes the form:
\begin{align}
G_{\bm i}^{-1} \check G_{\bm i \bm j}(\omega_m) = \delta_{\bm i \bm j}, \label{gorkov_eq_ml}
\end{align}
\begin{align}
G_{\bm i}^{-1} = \tau_z \left( \hat j + \mu_S - \check \Delta_{\bm i} i \sigma_y - \bm h \check {\bm \sigma}  \right) + i \omega_m .
\label{G_i}
\end{align}
where $\check \Delta_{\bm i} = \Delta_{\bm i} \tau_+ + \Delta_{\bm i}^* \tau_-$ with $\tau_\pm  = (\tau_x \pm i \tau_y)/2$ and $\check {\bm \sigma} = \bm \sigma (1+\tau_z)/2 + \bm \sigma^* (1-\tau_z)/2$ is the quasiparticle spin operator. Further we consider the Green's function in the mixed representation:
\begin{eqnarray}
\check G(\bm R, \bm p) = F(\check G_{\bm i \bm j}) = \int d^2 r e^{-i \bm p(\bm i - \bm j)}\check G_{\bm i \bm j},
\label{mixed}
\end{eqnarray}
where $\bm R=(\bm i+\bm j)/2$ and the integration is over $\bm i - \bm j$. The presence of a homogeneous supercurrent flowing in the $\mathrm{NbSe_2}$ layer is taken into account via the phase gradient of the superconducting order parameter $\check \Delta(\bm R) = \Delta \tau_x e^{i \bm q \bm R}$, where $\Delta$ is the absolute value of the order parameter, which does not depend on coordinates, and $\bm q$ is the superconducting phase gradient (total momentum of the Cooper pair) aligned with the supercurrent $\bm j_e$, which makes angle $\theta$ with the direction of $\bm h$. Making the gauge transformation
\begin{eqnarray}
\check {G}(\bm R, \bm p) = 
e^{i \bm q \bm R \tau_z/2}  \check G_q(\bm p)  
e^{-i \bm q \bm R \tau_z/2} ,
\label{unitary}
\end{eqnarray}
we turn to the Green's function $\check G_q(\bm p) $, which does not depend on the spatial coordinates. We also define the following  transformed Green's function to simplify further calculations and to present the Gor'kov equation in a more common form:
\begin{eqnarray}
\check {\tilde G}_q(\bm p) = 
\left(
\begin{array}{cc}
1 & 0 \\
0 & -i\sigma_y
\end{array}
\right)_\tau  \check G_q(\bm p)  
\left(
\begin{array}{cc}
1 & 0 \\
0 & -i\sigma_y
\end{array}
\right)_\tau ,
\label{unitary}
\end{eqnarray}
where subscript $\tau$ means that the explicit matrix structure corresponds to the particle-hole space. Then we obtain the following Gor'kov equation for $\check {\tilde G}_q(\bm p)$:
\begin{align}
    \left( 
    i\omega_m\tau_z+i\tau_y \check \Delta -\bm h\bm \sigma \tau_z
    \right) \check {\tilde G}_q(\bm p)+\Xi(\bm p)=1,
    \label{eq:Gorkov}
\end{align}
where introducing the explicit structure of the Green's function $\check {\tilde G}_q(\bm p)$ in the particle-hole space
\begin{align}
    \check {\tilde G}_q(\bm p)=\left(
\begin{array}{cc}
 G_q(\bm p) &  F_q(\bm p) \\
\bar F_q(\bm p) & \bar G_q(\bm p)
\end{array}
\right)_\tau,
\end{align}
we can write the term $\Xi(\bm p)$ as follows:
\begin{align}
    \Xi=\left(
\begin{array}{cc}
\hat \xi_S(\bm p)  G_q(\bm p)& \hat \xi_S(\bm p+\bm q/2)  F_q(\bm p)\\
\hat \xi_S(\bm p-\bm q/2) \bar F_q(\bm p)& \hat \xi_S(\bm p) \bar G_q(\bm p)
\end{array}
\right)_\tau.
\end{align}
Here $\hat \xi_{S}$ is a diagonal matrix in spin space, describing the normal state electron spectrum of the $\mathrm{NbSe_2}$, with elements $\xi_{S}^\sigma(\bm p)=\mu_S+\sum\limits_{<\bm 0\bm j>,\sigma} t_{S}^{\bm 0\bm j,\sigma}e^{i\bm p \bm j }$. 

The superconducting order parameter is calculated from the self-consistency equation
\begin{eqnarray}
\Delta =  \lambda  T \sum \limits_{\omega_m}\int \frac{d^2 p}{(2\pi)^2} \frac{{\rm Tr}[\check {\tilde G}_q\sigma_0(\tau_x-i\tau_y)]}{4} .
\label{SC}    
\end{eqnarray}
Further we work near the critical temperature and expand the Green's function $\check {\tilde G}_q(\bm p)$ in powers of the superconducting order parameter $\Delta$:
\begin{align}
     G_q(\bm p)= G_q^0(\bm p)+\Delta^2 G_q^2(\bm p)+..., \\ \nonumber
     \bar F_q(\bm p)=\Delta  \bar F_q^1(\bm p)+...~~~~~~
\end{align}
Then the solution of Eq.~(\ref{eq:Gorkov}) for the leading contribution $\bar F_q^1(\bm p)$ to the anomalous Green's function takes the form:
\begin{align}
    \bar F_q^1(\bm p)=(-i\omega_m+\bm h\bm \sigma-\hat \xi_S(\bm p-\bm q/2))^{-1} \times\\ \nonumber
    \times (i\omega_m-\bm h\bm \sigma-\hat \xi_S(\bm p))^{-1}.~~~~~~~~
\end{align}
For the leading superconducting contribution to the normal  Green's function one obtains:
\begin{align}
    G_q^2(\bm p)=-(i\omega_m-\bm h\bm \sigma-\hat \xi_S(\bm p))^{-1}\times \nonumber \\  
    (-i\omega_m+\bm h\bm \sigma-\hat \xi_S(\bm p-\bm q/2))^{-1} \times \nonumber \\ 
    \times (i\omega_m-\bm h\bm \sigma-\hat \xi_S(\bm p))^{-1}~~~~~~
    \label{eq:G_leading}
\end{align}
The electric $\bm j_e$ and spin $\bm j_s^i$ ($i=x,y,z$ means the $i$-th component in spin space) currents can be calculated from the matrix current $\hat {\bm j}$ with the following elements in spin space ($\alpha,\beta = \uparrow, \downarrow$): 
\begin{align}
    \bm j^{\alpha\beta}=T\sum_{\omega_m}\int \frac{d^2 p}{(2\pi)^2} \bm v_F^{\alpha\beta} G_{q,\alpha \beta} (\bm p)
    \label{eq:current_matrix}
\end{align}
where
\begin{align}
    \bm v_F^{\uparrow\uparrow,\downarrow\downarrow}(\bm p)=\frac{d\xi_S^{\uparrow,\downarrow}}{d\bm p}
\end{align}
and
\begin{align}
    \bm v_F^{\uparrow\downarrow,\downarrow\uparrow}(\bm p)=\frac{d}{d\bm p}(\sum\limits_{<\bm 0\bm j>,\sigma} \frac{t_{S}^{\bm 0\bm j,\sigma}+t_{S}^{\bm 0\bm j,-\sigma}}{2}e^{i\bm p \bm j }).
\end{align}
Then
\begin{align}
    \bm j_e=\bm j^{\uparrow\uparrow}+\bm j^{\downarrow\downarrow},~~~\bm j_s^i={\rm Tr}[\hat {\bm j} \sigma_i] .
    \label{eq:currents}
\end{align}
The electric and spin current are normalized to the value of the electric charge $e$ and the electron spin $\hbar/2$, respectively.

\subsection{Spin polarization of Cooper pairs}

\label{sec:field_spin}

In this subsection we present approximate analytical results for the spin structure, which is acquired by Cooper pairs  in the $\mathrm{NbSe_2}$ monolayer as a result of the simultaneous presence of the Ising spin-orbit interaction and the applied magnetic field. The Fermi surface of the $\mathrm{NbSe_2}$ monolayer is presented in Fig.~\ref{fig:fermi}(a). It consists of a central contour surrounding the $\Gamma$-point and six contours around $K$ and $K'$-points, which are spin-split due to the Ising-type spin orbit coupling \cite{Wickramaratne2020}. In this section we assume that the characteristic energy value of the spin splitting of the $K$-valleys $V_{SO}$ is much larger than any other energy scales in our system, $h$ and $\Delta$. This assumption is in good agreement with results of the DFT calculations for the $\mathrm{NbSe_2}$ monolayer, where the Ising spin splitting of the order of $100$meV was reported \cite{Wickramaratne2020}. For our analytical consideration we also neglect the spin splitting of the Fermi contours around the  $\Gamma$-point  because it is much smaller that the spin splitting around $K$-points. Under such assumptions we can integrate the anomalous Green's function $\bar F_q^1(\bm p)$ over the absolute value of the electron momentum in the vicinity of each of the Fermi contours (the so-called $\xi$-integration). As a result we obtain the anomalous Green's function 
\begin{align}
    \bar f_q^{1, \nu,s}(\bm n)=-\frac{1}{i\pi}\int d\xi \bar F_q^1(\bm p),
\end{align}
which only depends on the direction $\bm n = \bm p/p$ on the Fermi surface, but not on the absolute value of the electron momentum $p$. This is the same procedure that is used to go to the quasiclassical approximation in the theory of superconductivity\cite{Serene1983}. However, here we do not derive any quasiclassical equations for $\bar f_q^{1, \nu,s}(\bm n)$, since our system is spatially homogeneous. The $\xi$-integrated Green's function is determined at each of the Fermi surface contours around $K$-points, which are marked by the valley index $\nu = \pm 1$ and spin index $s=\pm 1$ [see Fig.~\ref{fig:fermi}(a)]. To obtain the $\xi$-integrated Green's function at the contour corresponding to spin $s$ and valley $\nu$ we should take 
\begin{align}
    \xi_S^s(\bm p)=\xi,~~~\xi_S^{-s}(\bm p)=s \nu V_{SO} .
\end{align}
Also expanding the normal state electron dispersion up to the leading order with respect to the condensate momentum $\bm q$ 
\begin{align}
    \xi_S^s(\bm p-\bm q/2)=\xi_S^s(\bm p)-{ \bm v}_F^{s,\nu} \bm q/2,
\end{align}
where ${\bm v}_F^{s,\nu} = \bm v_F^{ss}(\bm p = \bm p_F^{s,\nu})$ is the Fermi velocity at the corresponding Fermi-surface contour, we obtain the following expression for the $\xi$-integrated anomalous Green's function:
\begin{align}
    \bar f_q^{1, \nu,s}(\bm n)={\rm sgn}\omega_m\frac{-(\sigma_0+ s \sigma_z)/2+i((\bm z\times \bm h)\bm \sigma)\nu V_{SO}^{-1}}{i\omega_m-s(\bm z \bm h)-\bm v_F^{s, \nu}\bm q/4},
    \label{eq:f_quasiclassical}
\end{align}

The anomalous Green's function can be always represented in the form
\begin{align}
    \bar f_q^{1, \nu,s}(\bm n)=f_s^{\nu,s} (\bm n) + \bm d^{\nu,s} (\bm n) \bm \sigma ,
    \label{eq:d_vector}
\end{align}
where $f_s^{\nu,s}$ accounts for singlet superconducting correlations and $\bm d^{\nu,s}$ describes triplet correlations. The averaged spin of a Cooper pair  
\begin{align}
    \bm S \propto i \hbar T \sum \limits_{s,\nu,\omega_m} \int \frac{d \Omega}{2\pi} \bm d^{\nu,s *} \times \bm d^{\nu,s} .
    \label{eq:spin_total}
\end{align}
From Eqs.~(\ref{eq:f_quasiclassical}) and (\ref{eq:d_vector}) we obtain
\begin{align}
    \bm d^{\nu,s *} \times \bm d^{\nu,s} =\frac{-i s \nu V_{SO}^{-1}}{\omega_m^2+(s(\bm z \bm h)-\bm v_F^{s,\nu}\bm q/4)^2} (\bm z \times (\bm z \times \bm h)) .
    \label{eq:spin_pair}
\end{align}

It is worth noting that the non-unitary triplet pairing \cite{Leggett1975,Sigrist1991} arises here due to the fact that there are nonzero $\bm z$ and $\bm z \times \bm h$ components of the $\bm d^{\nu,s}$-vector with the $\pi/2$-shift between them. This does not occur if one considers the Ising-type spin-orbit coupling in the quasiclassical approximation, which implies small spin-orbit splitting of the Fermi surface contours around the $K$-points \cite{Mockli2020}. In this consideration we do not calculate the $\xi$-integrated anomalous Green's function at the Fermi-surface contour around the $\Gamma$-point since in the framework of our approximation the non-unitary triplet is absent there due to the absence of the Ising-type spin-orbit splitting. 

The total average spin of the Cooper pair is determined by the integration of $\bm d^{\nu,s *} \times \bm d^{\nu,s}$ over all the Brillouin zone, see Eq.~(\ref{eq:spin_total}). However, the pairs formed by electrons belonging to outer and inner (with respect to the corresponding $K$-points) Fermi contours contribute to the spin supercurrent with different condensate velocities, see the next subsection. For this reason it is useful to calculate the average spin of the Cooper pairs integrated over outer and inner Fermi surface contours separately. These quantities are denoted by $\bm S_{out}$ and $\bm S_{in}$ and take the following form:
\begin{align}
\bm S_{out(in)} \propto T\sum_{\omega_m} (\pm N_{out(in)})\frac{(\bm z \times (\bm z \times \bm h))}{(\omega_m^2+(\bm z \bm h)^2)V_{SO}}, 
\label{eq:spin_out_in}
\end{align}
where $N_{out(in)}$ is the normal state density of states at the outer (inner) Fermi surface contours. The sign $\pm$ originates from the factor $s \nu$, which is $+1(-1)$ for outer (inner) Fermi surface contours.   

\subsection{Analytical results for the spin current}

\label{sec:field_current_analytical}

As it follows from Eqs.~(\ref{eq:currents}), (\ref{eq:current_matrix}) and (\ref{eq:G_leading}), the spin supercurrent is determined by the $\xi$-integrated leading superconducting contribution to the normal Green's function:
\begin{align}
    g_q^{2, \nu,s}(\bm n)=-\frac{1}{i\pi}\int d\xi G_q^2(\bm p),
\end{align}
which after explicit $\xi$-integration of Eq.~(\ref{eq:G_leading}) takes the form:
\begin{align}
    g_q^{2, \nu,s}(\bm n)=-{\rm sgn}\omega_m \times  \nonumber  \\ \frac{(\sigma_0+s \sigma_z)+2i(\bm z\times(\bm z\times \bm h))\bm \sigma s \nu V_{SO}^{-1}}{4(i\omega_m-s(\bm z \bm h)-\bm v_F^{s, \nu}\bm q/4)^2}
\end{align}
After performing the remaining integration over the directions at the Fermi surface in Eq.~(\ref{eq:current_matrix}) and making use of Eq.~(\ref{eq:currents}) we obtain the only nonzero in spin space component of the spin current:
\begin{align}
    j_s^{\bm z\times(\bm z\times \bm h)}=\frac{\pi T\bm q |z\times(\bm z\times \bm h)|}{8V_{SO}} \times \nonumber \\
    \sum_{\omega_m}\frac{N_{out}v_{F,out}^{\uparrow\downarrow} v_{F,out}^{\uparrow\uparrow}-N_{in}v_{F,in}^{\uparrow\downarrow}v_{F,in}^{\downarrow\downarrow}}{(i\omega_m-(\bm h \bm z))^3}
    \label{eq:current_quasiclassical}
\end{align}

\begin{align}
    v_{F,out}^{\uparrow\uparrow(\downarrow\downarrow)}= |{\bm v}_F^{\uparrow(\downarrow),\nu=-1(+1)}|
\end{align}

\begin{align}
    v_{F,in}^{\uparrow\uparrow(\downarrow\downarrow)}= |{\bm v}_F^{\uparrow(\downarrow),\nu=+1(-1)}|
\end{align}

\begin{align}
    v_{F,out(in)}^{\uparrow\downarrow}=|\bm v_F^{\uparrow\downarrow}({\bm p=\bm p_F^{\uparrow,\nu=-1(+1)}})|
\end{align}
Comparing Eq.~(\ref{eq:current_quasiclassical}) to Eq.~(\ref{eq:spin_out_in}) we can conclude that it can be interpreted as a sum of two different components of the condensate of spinful Cooper pairs with spins $\bm S_{out(in)}$ belonging to outer and inner Fermi surface contours. Both flows are induced by the applied supercurrent with the condensate momentum $\bm q$. 

\subsection{Numerical results for the spin current}

\label{sec:field_current_numerical}

\begin{figure}[tb]
	\begin{center}
		\includegraphics[width=80mm]{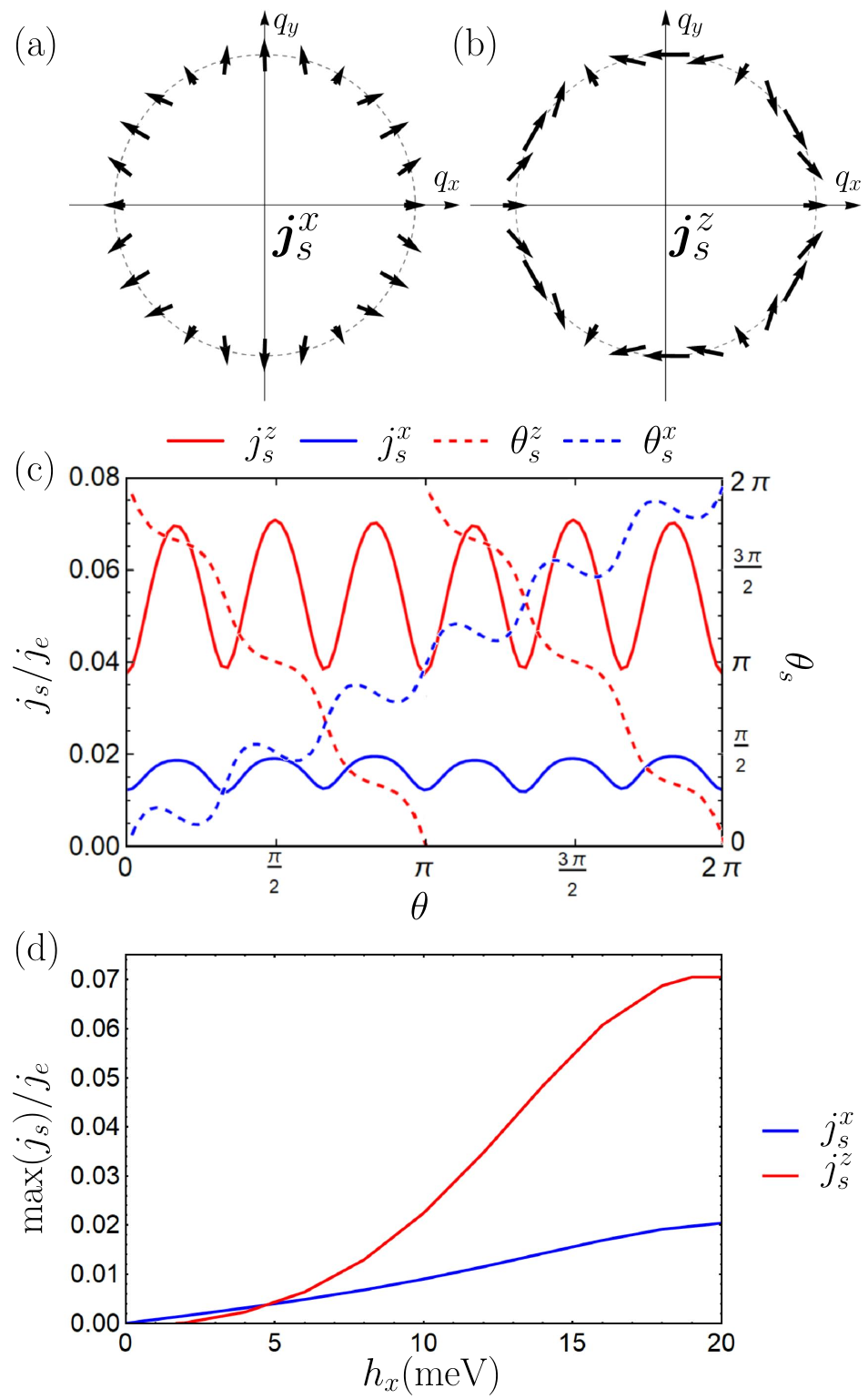}
\caption{Numerical results for the spin current. External magnetic field is applied in plane of the S layer along the $x$-axis, that is $\bm h = h_x \bm x$. (a) Amplitude and the direction of the black arrows show the amplitude and the direction of the spin current $\bm j_s^{\bm z \times (\bm z \times \bm h)} \equiv \bm j_s^x$ for a given direction of the condensate momentum $\bm q$ coinciding with the direction of the charge supercurrent $\bm j_e$. (b) The same as in (a) but for $\bm j_s^z$. (c) Amplitude of the spin current components $j_s^x$ (solid blue curve), $j_s^z$ (solid red) and angles $\theta_s^x$ (dashed blue), $\theta_s^z$ (dashed red) between the direction of the corresponding spin currents $\bm j_s^x$, $\bm j_s^z$ flows and the $x$-axis as functions of the angle $\theta$ between $\bm q$ and the $x$-axis at $h_x=20$meV. (d) Maximal (with respect to the direction of $\bm q$) value of the ratio $j_s/j_e$ for a given absolute value of the condensate momentum $q$ close to $q_c$ as a function of $h_x$. Red and blue lines correspond to two nonzero spin current components $j_s^z$ and $j_s^x$, respectively. Larger values of $h_x$ fully suppress superconductivity.}
 \label{fig:numerical}
	\end{center}
 \end{figure}

The analytical results for the spin current presented in the previous section are valid under the assumption $h/V_{SO} \ll 1$. In this limit the charge supercurrent generates a spin supercurrent flowing in the same direction and carrying spin directed along the vector $\bm z \times (\bm z \times \bm h)$. It means that if we apply the external magnetic field in plane of the $\mathrm{NbSe_2}$ the spin current carries the spin component aligned with $\bm h$. In this section we investigate the spin supercurrent numerically in the framework of the Bogolubov-de Gennes approach  beyond the assumption $h/V_{SO} \ll 1$. As it follows from results presented in this section  the analytical expressions obtained under the assumption $h/V_{SO} \ll 1$ work well only at $h/V_{SO} \lesssim 0.03$. Since superconductivity in a monolayer  $\mathrm{NbSe_2}$ is protected by the Ising-type spin-orbit coupling \cite{Xi2016,delaBarrera2018,Saito2016,Dvir2018,Sohn2018} and can survive up to rather strong in-plane field of the order of $0.05-0.1 ~ h/V_{SO}$, such a numerical analysis  is important.

We diagonalize Hamiltonian (\ref{eq:hamiltonian}) by the Bogoliubov transformation:
\begin{align}
c_{\bm i\sigma}=\int \frac{d^2 p}{(2\pi)^2}e^{i\bm p \bm i}\left(\sum\limits_n u_{n\sigma}(\bm p)\hat b_n+v^{*}_{n\sigma}(\bm p)\hat b_n^\dagger\right) , 
\label{bogolubov}
\end{align}
where $\hat b_n^\dagger(\hat b_n)$ are the creation (annihilation) operators of Bogoliubov quasiparticles. Then
the resulting Bogoliubov – de Gennes equations take the form:
\begin{align}
\xi_S^\sigma(\bm p+\bm q/2)u_{n,\sigma} + \sigma \Delta v_{n,-\sigma}+ \\ \nonumber
+(\bm h \bm{\sigma})_{\sigma\alpha}u_{n,\alpha} & = \varepsilon_n u_{n,\sigma} \nonumber \\  
-\xi_S^\sigma(\bm p-\bm q/2)v_{n,\sigma} + \sigma \Delta^* u_{n,-\sigma}+ \\ \nonumber
+(\bm h\bm{\sigma}^*)_{\sigma\alpha}v_{n,\alpha} & = -\varepsilon_n v_{n,\sigma}, 
\label{bdg}
\end{align}

\begin{align}
\Delta = \lambda \sum\limits_n \int \frac{d^2 p}{(2\pi)^2}(u_{n,\downarrow} v_{n,\uparrow}^{*}(1-f_n)+u_{n,\uparrow} v_{n,\downarrow}^{*}f_n),
\end{align}
where $f_n=1/[1+{\rm exp}(-\varepsilon_n/T)]$ is the Fermi distribution.

The matrix current can be calculated via the solutions of the Bogolubov-de Gennes equation as follows:
\begin{align}
    \bm j^{\alpha\beta}=\sum_{n}\int \frac{d^2 p}{(2\pi)^2} \bm v_F^{\alpha\beta} \left(u_{n\alpha} u_{n\beta}^{*}f_n+v_{n\alpha}^{*} v_{n\beta}(1-f_n)\right)
\end{align}

The spin current component $\bm j_s^{\bm z \times (\bm z \times \bm h)}$ calculated numerically in the framework of the Bogolubov-de Gennes approach is demonstrated in Figs.~\ref{fig:numerical}(a) and (c) in different representations. In Fig.~\ref{fig:numerical}(a)  the direction and the absolute value of $\bm j_s^{\bm z \times (\bm z \times \bm h)}$ for a given direction of the charge supercurrent $\bm j_e$ (or equivalently a given direction of the condensate momentum $\bm q$), is shown by black arrows. In Fig.~\ref{fig:numerical}(c) the amplitude of $\bm j_s^{\bm z \times (\bm z \times \bm h)}$ and the angle $\theta_s^{\bm z \times (\bm z \times \bm h)}$ between the direction of $\bm j_s^{\bm z \times (\bm z \times \bm h)}$ flow and the $x$-axis (along which the magnetic field is applied) are plotted as functions of the angle $\theta$ by solid and dashed blue lines, respectively. It is seen that $\bm j_s^{\bm z \times (\bm z \times \bm h)}(\bm q)$ has hexagonal symmetry, which originated from the hexagonal symmetry of the Brillouin zone. We can also see a slight deviation of the direction of the spin current $\bm j_s^{\bm z \times (\bm z \times \bm h)}$ flow from the direction of the charge current $\bm j_e$, which is directed strictly radially. It is worth noting that 
the perpendicular  component of the spin supercurrent, which appears in our numerical results, originates from the crystal structure of the superconductor. It does not possess a finite chirality and is not related to physical mechanisms of a spin Hall current generation, which would give a vector structure $\bm j_s \propto [\bm j_e \times \bm z]$.

Our numerical calculation also demonstrates that the spin current also transfers a non-zero spin projection along the $z$-axis $\bm j_s^z$. It is shown in Figs.~\ref{fig:numerical}(b) and (c). The symmetry of this component is reduced to a three-fold rotation axis. This is because this component is odd with respect to $\bm z \to - \bm z$, which is equivalent to the reversal of the Ising spin splitting. From Fig.~\ref{fig:fermi}(a) it is seen that accounting for the Ising spin splitting of the Fermi contours around $K$-points reduces the hexagonal rotational symmetry of the electronic structure to the three-fold rotation axis. $\bm j_s^z$ has much stronger component perpendicular to the charge current as compared to $\bm j_s^{\bm z \times (\bm z \times \bm h)}$. Very interesting property of $\bm j_s^z$ that it manifests a rectification effect.  $j_{s}^z$ conserves its sign under the sign reversal of the condensate momentum, that is $j_{s}^z \to j_{s}^z$ at $\bm q \to - \bm q$. The physical reason for this rectification effect is that the spin component of the Cooper pair spin directed along the $z$-axis is zero in the absence of the applied supercurrent. It is induced by the supercurrent, which leads to the fact that $\bm j_s^z \propto q^2$ at small values of the condensate momentum, as we have checked numerically.

In Fig.~\ref{fig:numerical}(d) we plot the maximal absolute value of the spin supercurrent (as its ratio to the amplitude of the charge supercurrent producing it) in dependence on the applied magnetic field $h_x$. At small $h_x$ $j_s^{\bm z \times (\bm z \times \bm h)} \equiv j_s^x \propto h_x/V_{so}$ in agreement with our analytical results. In this region of small $h_x$ $j_s^z \propto (h_x/V_{so})^2$, which is beyond the accuracy of the analytical approximation. However, from Fig.~\ref{fig:numerical}(d) we can see that the expressions obtained under the assumption $h/V_{so} \ll 1$ work well only at $h/V_{so} \lesssim 0.03$. Realistic magnetic fields achievable in laboratory conditions correspond to $h/V_{so} \lesssim 10^{-2}$. That is, with good accuracy they fall within the range of parameters where our analytical description is valid. However, in Fig. 3 we present numerical results for a much larger range of $h_x$ because effective exchange fields of this scale can be achieved in S/F heterostructures, see Sec.~\ref{heterostructure}.  Beyond the region $h/V_{so} \ll 1$ the spin current component $j_s^z$, which grows quadratically with $h_x$, becomes the most important contribution to the spin current. The value of the ratio $j_s/j_e$ also depends on $q$. Results presented in Fig.~\ref{fig:numerical}(d) are calculated at $q \approx q_c$, where $q_c$ is  the condensate momentum  corresponding to the critical supercurrent. Maximal possible values of the ratio $j_s/j_e$, which can be considered as efficiency of the charge to spin current conversion in our system, can be achived at $q \to q_c$ and according to our calculations can be estimated $\sim 0.1$.

It is worth noting that here we only consider equilibrium dissipationless transport of charge and spin carried by Cooper pairs. A quasiparticle current is absent. In this case the temperature dependence of the spin current is only determined by the BCS-like temperature dependence of the superconducting order parameter.

\section{$\mathrm{NbSe_2}/\mathrm{VSe_2}$ heterostructure}
\label{heterostructure}

\begin{figure}[tb]
	\begin{center}
		\includegraphics[width=70mm]{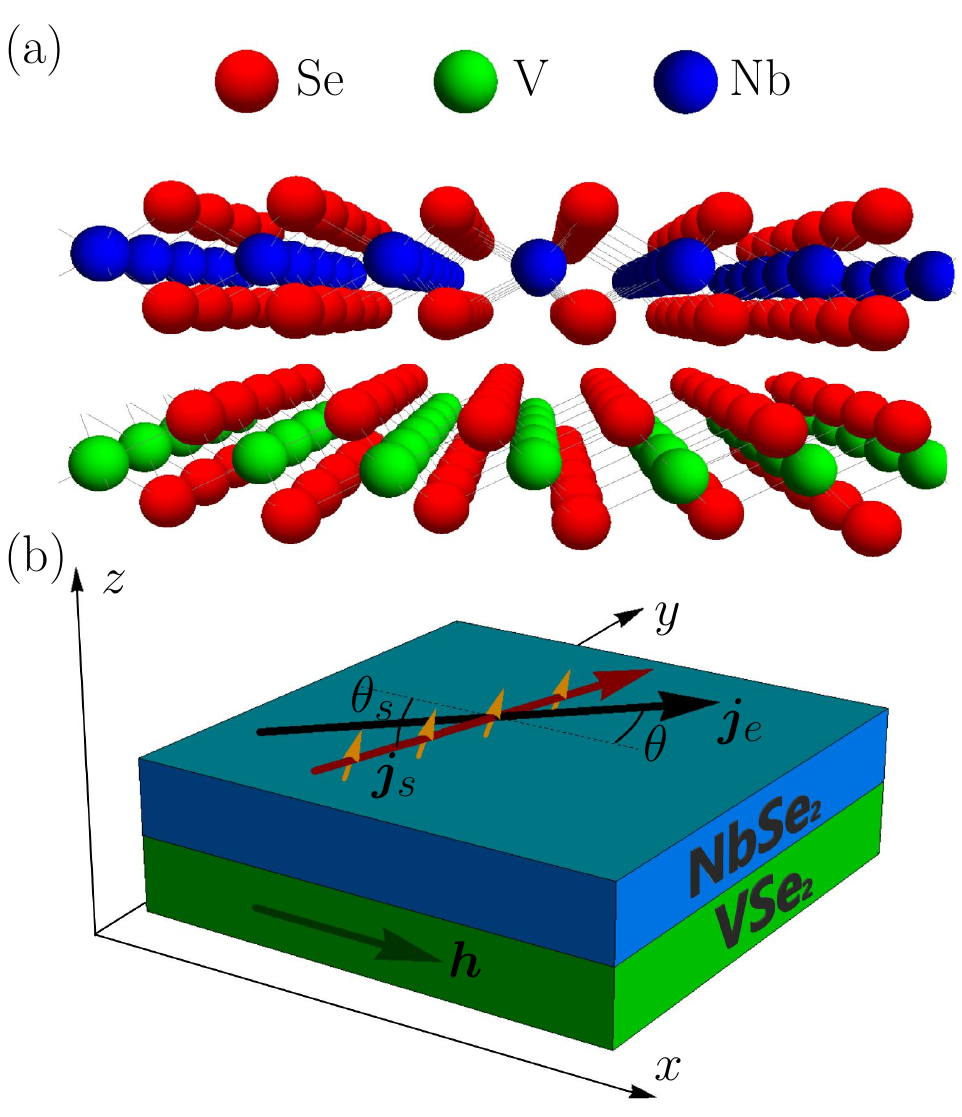}
\caption{(a) Atomic structure of the bilayer; (b) sketch of the heterostructure and visualization of the mutual directions of the internal exchange field $\bm h$ in $\mathrm{VSe_2}$, applied supercurrent $\bm j_e$ (black arrow) and induced spin current $\bm j_s$ (red arrow for the spatial flow and yellow arrows for the carried spin).}
 \label{fig:sketch_heterostructure}
	\end{center}
 \end{figure}

Here we present results for the superconducting spin current, which is generated by the charge supercurrent in S/F heterostructures. As an example, we consider a heterostructure consisting of a monolayer superconducting $\mathrm{NbSe_2}$ and a monolayer ferromagnetic $\mathrm{VSe_2}$ \cite{Ma2012,Bonilla2018,Yu2019,Chua2020,Huang2023}. The choice of specific materials is not the only possible one and is due to the fact that the electronic spectra of a given heterostructure have already been calculated earlier \cite{Bobkov2024_vdW}. See Fig.~\ref{fig:sketch_heterostructure}(a) for detailed atomic structure of the bilayer and Fig.~\ref{fig:sketch_heterostructure}(b) for a sketch of the heterostructure and visualization of the mutual directions of the internal exchange field $\bm h$ in $\mathrm{VSe_2}$, applied supercurrent $\bm j_e$ and induced spin current $\bm j_s$.

\subsection{Model}

\label{sec:heterostructure_model}

The tight-binding single-band Hamiltonian, which reasonably fits the electronic spectrum of a given heterostructure in the vicinity of the Fermi-surface takes the form \cite{Bobkov2024_vdW}:
\begin{widetext}
\begin{eqnarray}
\hat H = \sum\limits_{\bm i,\alpha,\beta}\hat c^\dagger_{\bm i,\alpha}\left(\begin{matrix}0&0\\0&(\bm h\bm\sigma)_{\alpha,\beta}\\\end{matrix}\right)\hat c_{\bm i,\beta}-
\sum\limits_{\bm i,\sigma}\hat c^\dagger_{\bm i,\sigma}\left(\begin{matrix}\mu_S&0\\0&\mu_F\\\end{matrix}\right)\hat c_{\bm i,\sigma}+~~~~~~~~~~~~~~~~~~~~\nonumber \\ 
+\sum\limits_{\bm i}\left[\hat c_{\bm i,\uparrow}\left(\begin{matrix}\Delta&0\\0&0\\\end{matrix}\right)\hat c_{\bm i,\downarrow}+H.c.\right]- \sum\limits_{<\bm i \bm j>,\sigma}\hat c^\dagger_{\bm i,\sigma}\left(\begin{matrix}t_S^{\bm i \bm j,\sigma}&0\\0&t_F^{\bm i \bm j,\sigma}\\\end{matrix}\right)\hat c_{\bm j,\sigma}-\sum\limits_{\bm i,\sigma}\hat c^\dagger_{\bm i,\sigma}\left(\begin{matrix}0&t_{SF}\\t_{SF}&0\\\end{matrix}\right)\hat c_{\bm i,\sigma},~~~~~~~
\label{eq:hamiltonian_heterostructure}
\end{eqnarray}
\end{widetext}
where $\hat c_{\bm i,\sigma} = (c_{\bm i,\sigma}^S, c_{\bm i,\sigma}^F)^T$ is a vector composed of annihilation operators for electrons belonging to the S and F layers at site $\bm i$ in plane of each layer and for spin $\sigma = \uparrow, \downarrow$. $t^{\bm i \bm j,\sigma}_S,t^{\bm i \bm j,\sigma}_F$ are complex hopping elements in the S and the F layers, respectively. $\mu_{S,F}$ are chemical potentials of the S and F layers, respectively. $t_{SF}$ is the hopping element between the S and F layers. $\bm h$ is the exchange field of the F layer. We assume that the magnetization of the $\mathrm{VSe_2}$ layer and, correspondingly, $\bm h$ is in its plane (IP-configuration). The parameters extracted from the fits of the DFT data \cite{Bobkov2024_vdW} for $\mathrm{VSe_2}$ are listed in Table~\ref{tab:hopping_par_F}. The interlayer hopping was estimated to be $t_{SF} = 30$ meV from the DFT spectra of the $\mathrm{NbSe_2}/\mathrm{VSe_2}$ heterostructure \cite{Bobkov2024_vdW}. 

\begin{table}
\begin{center}
\begin{tabular}{|c|c|c|c|c|c|c|c|c|}
\hline
 $\mu_F$ & $t_F^0$ & $t_F^1$ & $t_F^2$ & $t_F^3$ & $t_F^4$ & $t_F^5$ & $\varphi_F$ & $h$ \\
\hline
 -18.8 & -22.2 & 93.4 & -65.4 & 17.3 & -23.6 & 8.1 & 0.2 & 401 \\
\hline
\end{tabular}
\end{center}
\caption{\label{tab:hopping_par_F}Parameters of the one-band tight-binding model fitted to the DFT-calculated electron spectra of $\mathrm {VSe_2}$. All values of the hopping amplitudes and other energies are given in meV.}
\end{table} 

We diagonalize Hamiltonian (\ref{eq:hamiltonian}) by the Bogoliubov transformation:
\begin{align}
c_{\bm i\sigma}^\eta=\int \frac{d^2 p}{(2\pi)^2}e^{i\bm p \bm i}\left(\sum\limits_n u_{n\sigma}^\eta(\bm p)\hat b_n+v^{\eta *}_{n\sigma}(\bm p)\hat b_n^\dagger \right) , 
\label{bogolubov}
\end{align}
where $\hat b_n^\dagger(\hat b_n)$ are the creation (annihilation) operators of the Bogoliubov quasiparticles. $\eta=S,F$ is the layer index. Then the resulting Bogoliubov – de Gennes equations take the form:
\begin{align}
\xi_\eta^\sigma(\bm p+\bm q/2)u_{n,\sigma}^{ \eta} + \sigma \Delta_{\eta} v^{ \eta}_{n,-\sigma}-t_{SF}u_{n,\sigma}^{\bar \eta}+ \\ \nonumber
+(\bm h_\eta \bm{\sigma})_{\sigma\alpha}u_{n,\alpha}^{ \eta} & = \varepsilon_n u_{n,\sigma}^{ \eta} \nonumber \\  
-\xi_\eta^\sigma(\bm p-\bm q/2)v_{n,\sigma}^{ \eta} + \sigma \Delta_{ \eta}^* u^{ \eta}_{n,-\sigma}-t_{SF}v_{n,\sigma}^{\bar \eta}+ \\ \nonumber
+(\bm h_\eta\bm{\sigma}^*)_{\sigma\alpha}v_{n,\alpha}^{ \eta} & = -\varepsilon_n v_{n,\sigma}^{ \eta}, 
\label{bdg}
\end{align}
where $\bar S,\bar F=F,S$, $\Delta_\eta=(\Delta, 0)^T$, $\bm h_\eta=(0, \bm h)^T$ are the superconducting order parameter, which is nonzero only in the S layer, and the exchange field, which exist only in the F layer, respectively. 

\begin{align}
\Delta_{S}= \lambda \sum\limits_n (u_{n,\downarrow}^{S} v_{n,\uparrow}^{S*}(1-f_n)+u_{n,\uparrow}^{S} v_{n,\downarrow}^{S*}f_n).
\end{align}

The matrix current can be calculated via the solutions of the Bogolubov-de Gennes equation as follows:
\begin{align}
    \bm j^{\alpha\beta}=\sum_{n,\eta}\int \frac{d^2\bm p}{(2\pi)^2} \bm v_F^{\alpha\beta} \left(u_{n\alpha}^{ \eta} u_{n\beta}^{\eta*}f_n+v_{n\alpha}^{ \eta*} v_{n\beta}^{ \eta}(1-f_n)\right)
\end{align}

\subsection{Gate-controllable spin current}

\label{sec:heterostructure_current}

\begin{figure}[tb]
	\begin{center}
		\includegraphics[width=80mm]{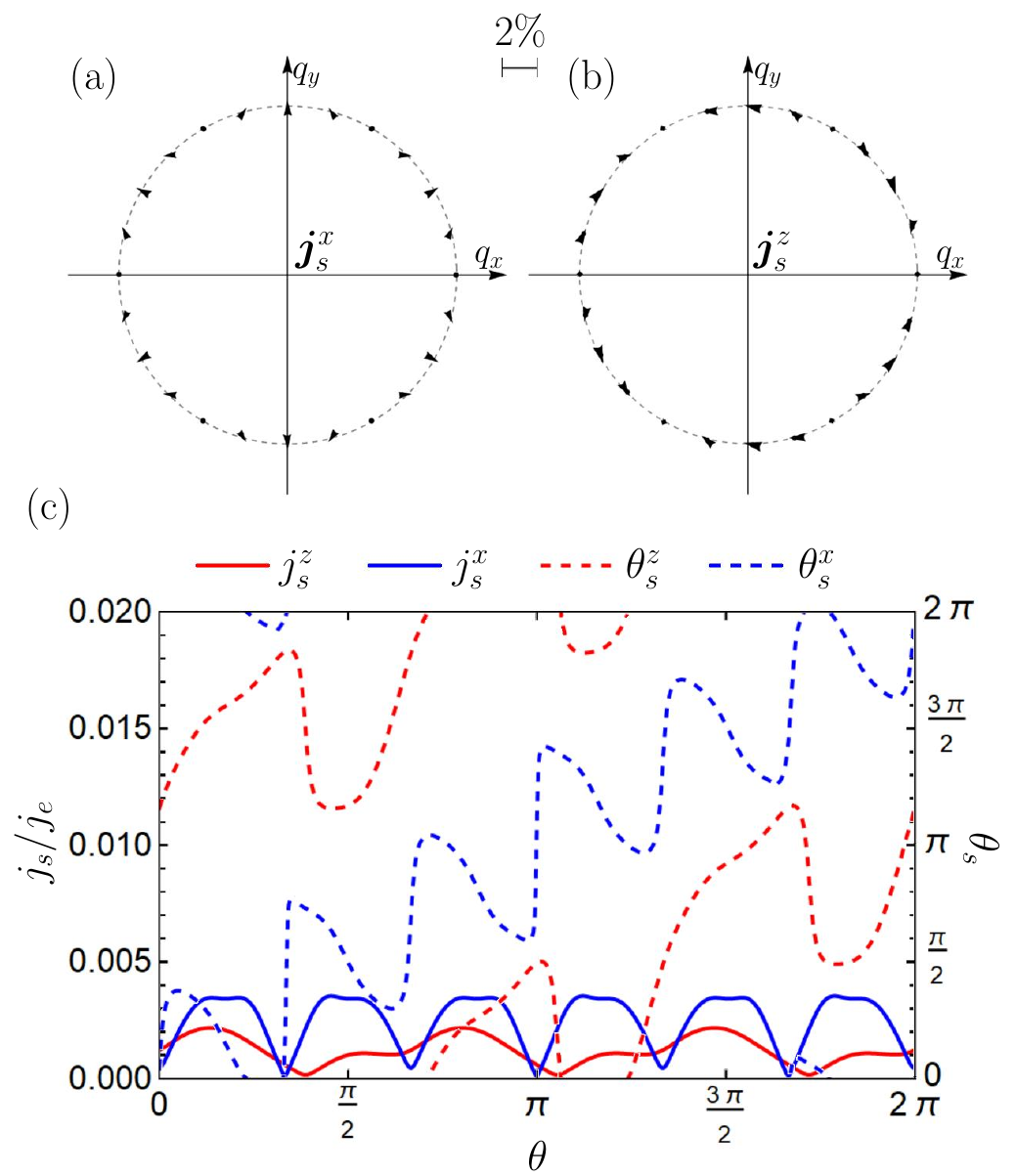}
\caption{Numerical results for the spin current in $\mathrm{NbSe_2}/\mathrm{VSe_2}$ heterostructure at no gating $V=0$. Magnetization of the $\mathrm{VSe_2}$ is in plane of the $\mathrm{NbSe_2}/\mathrm{VSe_2}$ interface, along the $x$-axis.  (a)-(b) Amplitude and direction
of the black arrows show the amplitude and the direction of
the spin currents $j_s^x$ (a) and $j_s^z$ (b) for a given direction of the condensate momentum $\bm q$, as in Fig.~\ref{fig:numerical}. The amplitude scale of the arrows is indicated in the upper part of the figure.  (c) Amplitudes of $\bm j_s^{x,z}$ and angles $\theta_s^{x,z}$ between the direction of $\bm j_s^{x,z}$ flow and the $x$-axis as functions of $\theta$. $t_{SF} = 30$meV.}
\label{fig:spin_current_heterostructure1}
 	\end{center}
  \end{figure}

\begin{figure}[tb]
	\begin{center}
		\includegraphics[width=80mm]{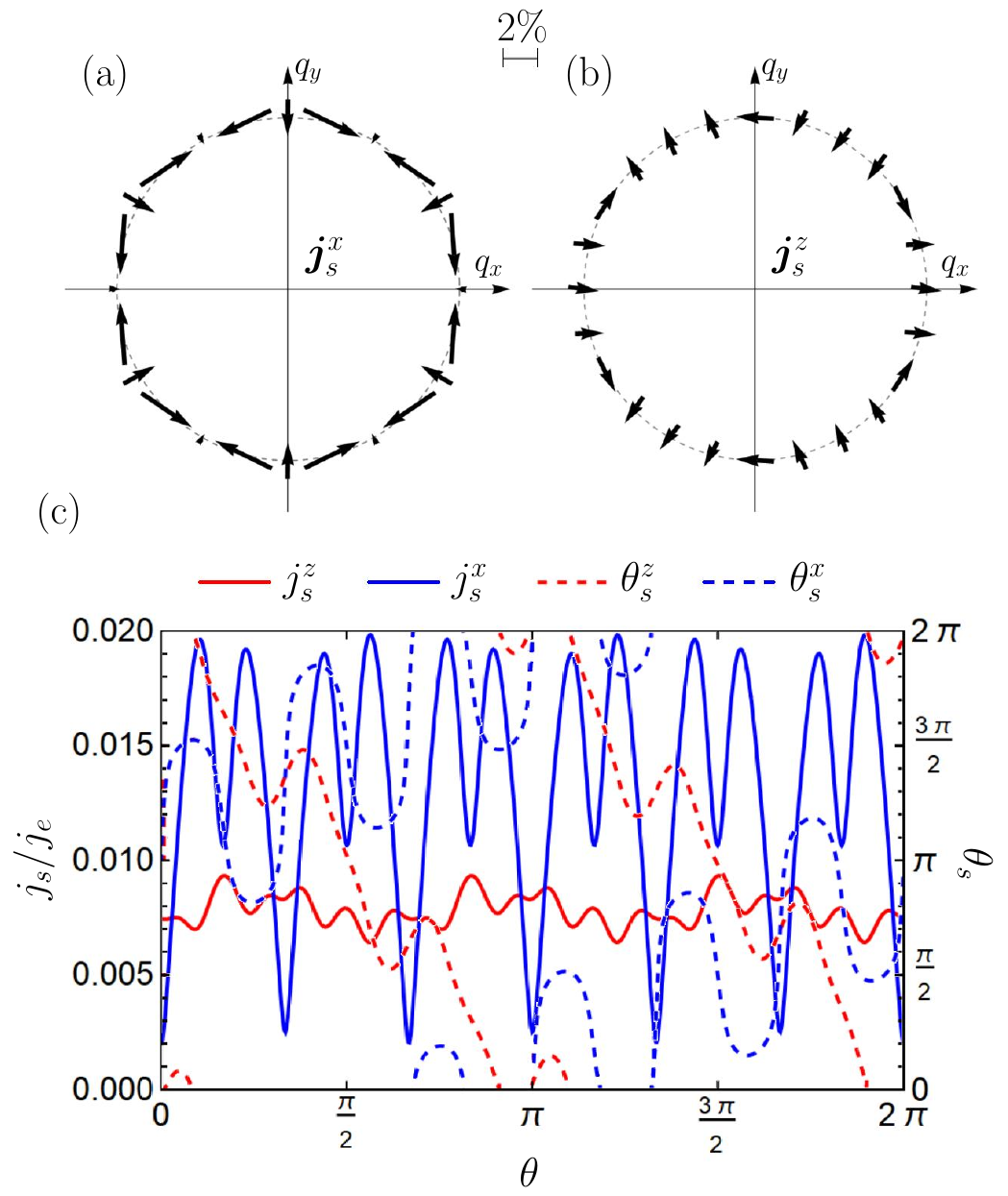}
\caption{Same as in Fig.~\ref{fig:spin_current_heterostructure1}, but for $V=540$meV.}
\label{fig:spin_current_heterostructure2}
 	\end{center}
  \end{figure}

\begin{figure}[tb]
	\begin{center}
		\includegraphics[width=80mm]{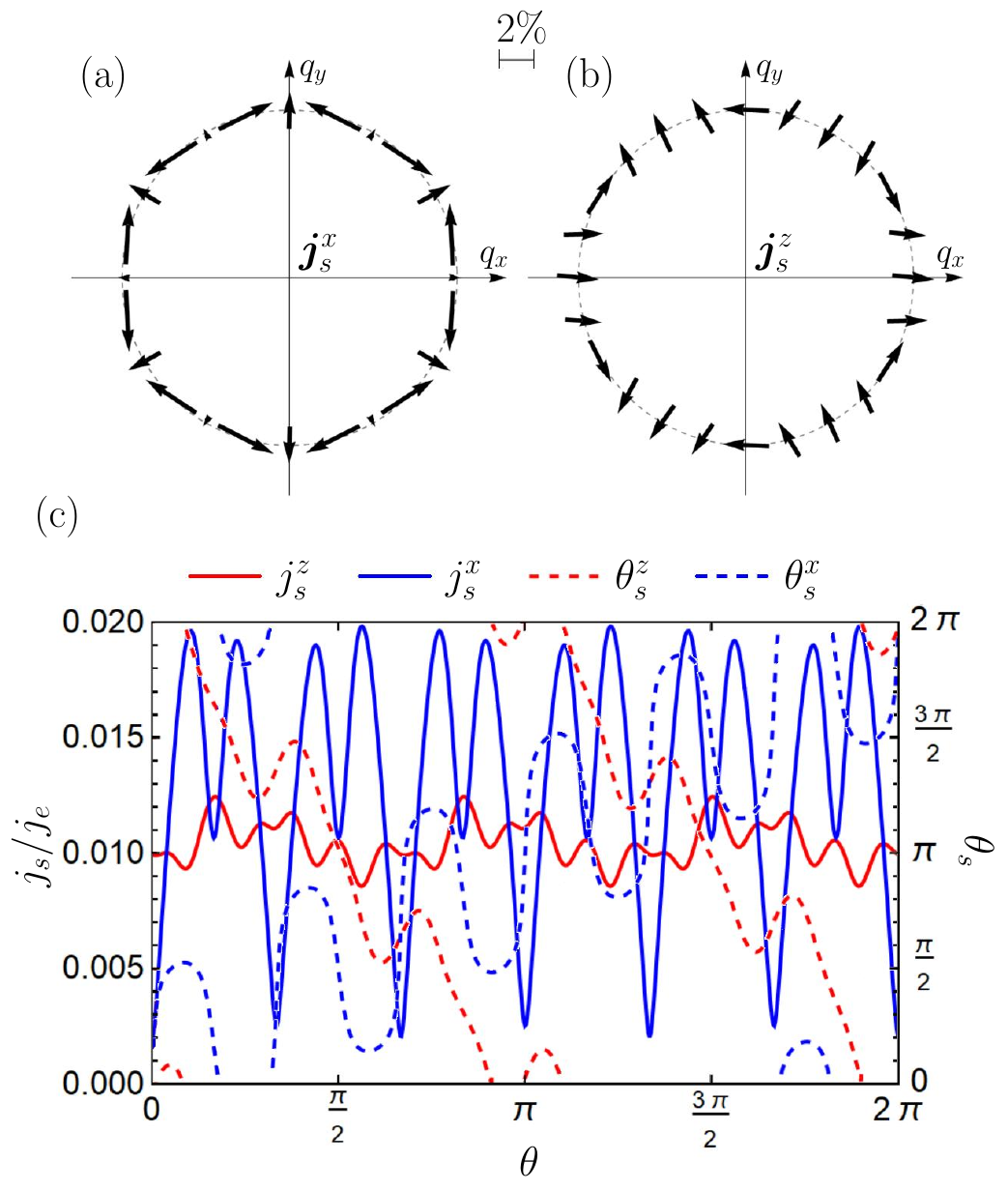}
\caption{Same as in Fig.~\ref{fig:spin_current_heterostructure1}, but for $V=-260$meV.}
\label{fig:spin_current_heterostructure3}
 	\end{center}
  \end{figure}

\begin{figure}[tb]
	\begin{center}
		\includegraphics[width=85mm]{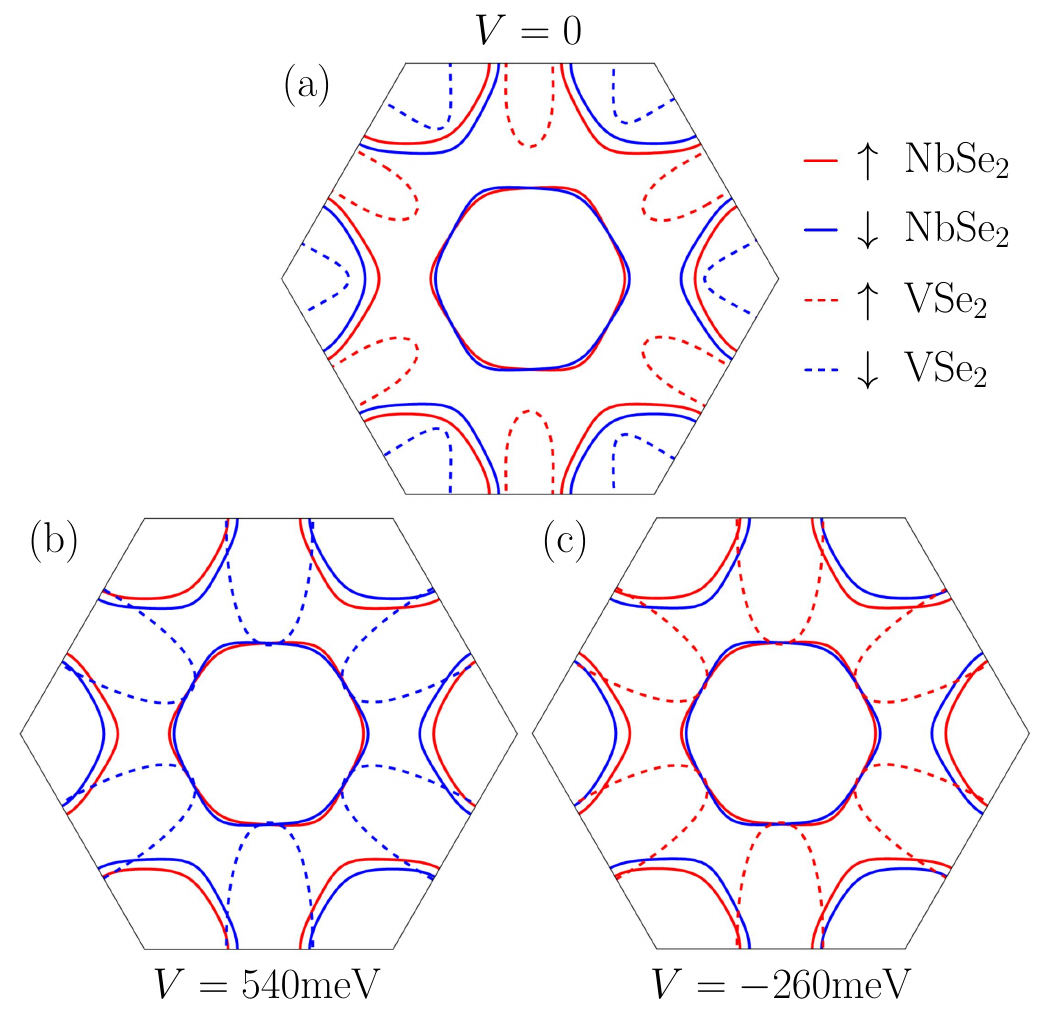}
\caption{Spin-split Fermi surfaces obtained from  single-band tight-binding Hamiltonians of the $\mathrm{NbSe_2}$ (solid) and $\mathrm{VSe_2}$ (dashed) monolayers at $t_{SF}=0$ with no gating potential in $\mathrm{VSe_2}$ (a); gating potential $V=540$ (b) and $V=-260$meV (c). Fermi surfaces belonging to the $\mathrm{NbSe_2}$ layer are shown by solid red (spin up) and solid blue (spin down) curves, and Fermi surfaces of the $\mathrm{VSe_2}$ layer are shown by dashed red (spin up) and dashed blue (spin down) curves.}
\label{fig:Fermi_heterostructure}
 	\end{center}
  \end{figure}

The superspin current for the $\mathrm{NbSe_2}/\mathrm{VSe_2}$  heterostructure is presented in Figs.~\ref{fig:spin_current_heterostructure1}-\ref{fig:spin_current_heterostructure3}. The results shown in these figures demonstrate that the qualitative physics discussed for the $\mathrm{NbSe_2}$ monolayer under the applied magnetic field survives here. Namely, we observe the same nonzero components of the spin current and $\bm j_s^z$ manifests the same rectification property. One of advantages of the heterostructure is apparently that in the case of a high-quality interface between the layers, according to the results of our calculations, it is possible to achieve sufficiently large values of the effective exchange field induced in the S layer by proximity effect. In turn, this makes it possible to achieve significantly higher ratios $j_s/j_e$ than is possible for realistic values of the magnetic fields applied to the isolated monolayer superconductor. 

However, the key result that distinguishes a heterostructure from a monolayer superconductor under the applied field is that the amplitude and sign of the spin current can be controlled by the gate voltage.  We apply the gating potential $V$ to the F layer. Amplitude (more precisely the ratio $j_s/j_e$) and direction of the spin currents $\bm j_s^x$ and $\bm j_s^z$ at zero gating $V=0$ are shown in Figs.~\ref{fig:spin_current_heterostructure1}(a) and (b), respectively. Fig.~\ref{fig:spin_current_heterostructure1}(c) represents the same data in the different form: the amplitudes of $\bm j_s^{x,z}$ and the angle $\theta_s^{x,z}$ between the direction of $\bm j_s^{x,z}$ flow and the $x$-axis are plotted as functions of the angle $\theta$.  It is seen that in this case the ratio $j_s/j_e$ is typically less than $10^{-2}$. At the same time from Figs.~\ref{fig:spin_current_heterostructure2} and \ref{fig:spin_current_heterostructure3}, which present the same information at $V=540$meV and $V=-260$meV, respectively, it is seen that the amplitude of the spin current can be strongly enhanced by gating. From  comparison of these figures we can see an even more striking result - the sign of the spin current can be changed by changing the gating potential. The reason for the observed dependence of the amplitude and sign of the spin currrent on the gating potential can be understood from Fig.~\ref{fig:Fermi_heterostructure}. In this Figure we demonstrate the Fermi surfaces of the heterostructure with no gating $V=0$ [Fig.~\ref{fig:Fermi_heterostructure}(a)] and at two different gating potentials, corresponding to Figs.~\ref{fig:spin_current_heterostructure2} and \ref{fig:spin_current_heterostructure3}, respectively. The plotted Fermi surfaces are calculated at $t_{SF}=0$ (separate S and F layers).  In Fig.~\ref{fig:Fermi_heterostructure}(a) it is seen that at $V=0$ the Fermi surfaces originating from the F (dashed curves) and S layers (solid curves around $K$-points) are too far from each other. It is a manifestation of the fact that the corresponding branches of the electronic spectra are also too far from each other and cannot be effectively hybridized at $t_{SF}=30$ meV. This distance between the $\mathrm{NbSe_2}$ and $\mathrm{VSe_2}$ electronic branches is of the order of $100$meV \cite{Bobkov2024_vdW}. For this reason at nonzero interlayer hopping $t_{SF} \lesssim 30$meV the effective exchange field, which is induced in the S layer due to the hybridization between the $\mathrm{VSe_2}$ and $\mathrm{NbSe_2}$ electronic spectra \cite{Bobkov2024_vdW}, is also weak.  On the contrary, in Figs.~\ref{fig:Fermi_heterostructure}(b)-(c) one can see that the modified by the gating potential Fermi surfaces of the $\mathrm{VSe_2}$ monolayer intersect with the  $\mathrm{NbSe_2}$ Fermi surfaces around the $K$-points. It leads to their strong hybridization at nonzero $t_{SF}$ and, consequently, to appearance of the higher effective exchange field in the S layer and a larger value of the spin supercurrent. Moreover, from a comparison of Figs.\ref{fig:Fermi_heterostructure}(b) and (c), one can see that by applying a gating potential of different signs, one can force the Fermi surfaces of $\mathrm{NbSe_2}$ to intersect with the Fermi surfaces of $\mathrm{VSe_2}$, corresponding to different spins. In its turn, it results in different signs of the effective exchange field induced in the $\mathrm{NbSe_2}$ by the hybridization of the electronic spectra, which leads to different signs of the spin current observed in Figs.~\ref{fig:spin_current_heterostructure2} and \ref{fig:spin_current_heterostructure3}, respectively. The results presented in Figs.~\ref{fig:spin_current_heterostructure2} and \ref{fig:spin_current_heterostructure3} correspond to the values of the gating potentials giving rise to a maximal degree of the hybridization between the electronic spectra of $\mathrm{NbSe_2}$ and $\mathrm{VSe_2}$ and, consequently, to approximately maximal possible values of the effective exchange field induced in the superconductor by proximity to the ferromagnet. Shifting the gate voltage results in weakening of the hybridization and a decrease of the spin current amplitude. 

\section{Conclusions}
\label{conclusions}

We develop a theory of dissipationless spin transport carried by a supercurrent in a monolayer vdW superconductor due to the simultaneous presence of the Ising-type spin-orbit coupling and a Zeeman field directed in plane of the layer. We consider two different possibilities: the Zeeman field  created by an external magnetic field and the Zeeman field generated by proximity effect with a ferromagnetic layer using the example of a bilayer heterostructure $\mathrm{NbSe_2}/\mathrm{VSe_2}$. We analyze the spin structure of triplet Cooper pairs and conclude that the simultaneous presence of the Ising-type spin-orbit coupling and the Zeeman field causes a partial conversion of the singlet pairing into the non-unitary triplet pairing with the averaged pair spin $\bm S \propto \bm z \times (\bm z \times \bm h)$. If the supercurrent is applied to the system, such pairs carry a superspin current. Furthermore, the triplet pairs acquire an additional spin component $\bm S_z \propto \bm z$ induced by the applied supercurrent and proportional to the condensate momentum, which also participate in the spin transport. For this spin current component we predict a rectification effect unlike the charge supercurrent that generates it. For the case of S/F bilayer heterostructures it is shown that the value and sign of
the spin polarization carried by the spin current can be controlled by gating.

\begin{acknowledgments}
The financial support from the Russian
Science Foundation via the RSF project No.24-22-00186 is acknowledged.  
\end{acknowledgments}

\bibliography{vdW_spin}

\begin{thebibliography}{86}%
\makeatletter
\providecommand \@ifxundefined [1]{%
 \@ifx{#1\undefined}
}%
\providecommand \@ifnum [1]{%
 \ifnum #1\expandafter \@firstoftwo
 \else \expandafter \@secondoftwo
 \fi
}%
\providecommand \@ifx [1]{%
 \ifx #1\expandafter \@firstoftwo
 \else \expandafter \@secondoftwo
 \fi
}%
\providecommand \natexlab [1]{#1}%
\providecommand \enquote  [1]{``#1''}%
\providecommand \bibnamefont  [1]{#1}%
\providecommand \bibfnamefont [1]{#1}%
\providecommand \citenamefont [1]{#1}%
\providecommand \href@noop [0]{\@secondoftwo}%
\providecommand \href [0]{\begingroup \@sanitize@url \@href}%
\providecommand \@href[1]{\@@startlink{#1}\@@href}%
\providecommand \@@href[1]{\endgroup#1\@@endlink}%
\providecommand \@sanitize@url [0]{\catcode `\\12\catcode `\$12\catcode `\&12\catcode `\#12\catcode `\^12\catcode `\_12\catcode `\%12\relax}%
\providecommand \@@startlink[1]{}%
\providecommand \@@endlink[0]{}%
\providecommand \url  [0]{\begingroup\@sanitize@url \@url }%
\providecommand \@url [1]{\endgroup\@href {#1}{\urlprefix }}%
\providecommand \urlprefix  [0]{URL }%
\providecommand \Eprint [0]{\href }%
\providecommand \doibase [0]{https://doi.org/}%
\providecommand \selectlanguage [0]{\@gobble}%
\providecommand \bibinfo  [0]{\@secondoftwo}%
\providecommand \bibfield  [0]{\@secondoftwo}%
\providecommand \translation [1]{[#1]}%
\providecommand \BibitemOpen [0]{}%
\providecommand \bibitemStop [0]{}%
\providecommand \bibitemNoStop [0]{.\EOS\space}%
\providecommand \EOS [0]{\spacefactor3000\relax}%
\providecommand \BibitemShut  [1]{\csname bibitem#1\endcsname}%
\let\auto@bib@innerbib\@empty
\bibitem [{\citenamefont {Datta}\ and\ \citenamefont {Das}(1990)}]{Datta1990}%
  \BibitemOpen
  \bibfield  {author} {\bibinfo {author} {\bibfnamefont {S.}~\bibnamefont {Datta}}\ and\ \bibinfo {author} {\bibfnamefont {B.}~\bibnamefont {Das}},\ }\bibfield  {title} {\bibinfo {title} {Electronic analog of the electro‐optic modulator},\ }\href {https://doi.org/10.1063/1.102730} {\bibfield  {journal} {\bibinfo  {journal} {Applied Physics Letters}\ }\textbf {\bibinfo {volume} {56}},\ \bibinfo {pages} {665} (\bibinfo {year} {1990})}\BibitemShut {NoStop}%
\bibitem [{\citenamefont {Gardelis}\ \emph {et~al.}(1999)\citenamefont {Gardelis}, \citenamefont {Smith}, \citenamefont {Barnes}, \citenamefont {Linfield},\ and\ \citenamefont {Ritchie}}]{Gardelis99}%
  \BibitemOpen
  \bibfield  {author} {\bibinfo {author} {\bibfnamefont {S.}~\bibnamefont {Gardelis}}, \bibinfo {author} {\bibfnamefont {C.~G.}\ \bibnamefont {Smith}}, \bibinfo {author} {\bibfnamefont {C.~H.~W.}\ \bibnamefont {Barnes}}, \bibinfo {author} {\bibfnamefont {E.~H.}\ \bibnamefont {Linfield}},\ and\ \bibinfo {author} {\bibfnamefont {D.~A.}\ \bibnamefont {Ritchie}},\ }\bibfield  {title} {\bibinfo {title} {Spin-valve effects in a semiconductor field-effect transistor: A spintronic device},\ }\href {https://doi.org/10.1103/PhysRevB.60.7764} {\bibfield  {journal} {\bibinfo  {journal} {Phys. Rev. B}\ }\textbf {\bibinfo {volume} {60}},\ \bibinfo {pages} {7764} (\bibinfo {year} {1999})}\BibitemShut {NoStop}%
\bibitem [{\citenamefont {Schmidt}\ \emph {et~al.}(2000)\citenamefont {Schmidt}, \citenamefont {Ferrand}, \citenamefont {Molenkamp}, \citenamefont {Filip},\ and\ \citenamefont {van Wees}}]{Schmidt00}%
  \BibitemOpen
  \bibfield  {author} {\bibinfo {author} {\bibfnamefont {G.}~\bibnamefont {Schmidt}}, \bibinfo {author} {\bibfnamefont {D.}~\bibnamefont {Ferrand}}, \bibinfo {author} {\bibfnamefont {L.~W.}\ \bibnamefont {Molenkamp}}, \bibinfo {author} {\bibfnamefont {A.~T.}\ \bibnamefont {Filip}},\ and\ \bibinfo {author} {\bibfnamefont {B.~J.}\ \bibnamefont {van Wees}},\ }\bibfield  {title} {\bibinfo {title} {Fundamental obstacle for electrical spin injection from a ferromagnetic metal into a diffusive semiconductor},\ }\href {https://doi.org/10.1103/PhysRevB.62.R4790} {\bibfield  {journal} {\bibinfo  {journal} {Phys. Rev. B}\ }\textbf {\bibinfo {volume} {62}},\ \bibinfo {pages} {R4790} (\bibinfo {year} {2000})}\BibitemShut {NoStop}%
\bibitem [{\citenamefont {Saitoh}\ \emph {et~al.}(2006)\citenamefont {Saitoh}, \citenamefont {Ueda}, \citenamefont {Miyajima},\ and\ \citenamefont {Tatara}}]{Saitoh2006}%
  \BibitemOpen
  \bibfield  {author} {\bibinfo {author} {\bibfnamefont {E.}~\bibnamefont {Saitoh}}, \bibinfo {author} {\bibfnamefont {M.}~\bibnamefont {Ueda}}, \bibinfo {author} {\bibfnamefont {H.}~\bibnamefont {Miyajima}},\ and\ \bibinfo {author} {\bibfnamefont {G.}~\bibnamefont {Tatara}},\ }\bibfield  {title} {\bibinfo {title} {Conversion of spin current into charge current at room temperature: Inverse spin-hall effect},\ }\href {https://doi.org/10.1063/1.2199473} {\bibfield  {journal} {\bibinfo  {journal} {Applied Physics Letters}\ }\textbf {\bibinfo {volume} {88}},\ \bibinfo {pages} {182509} (\bibinfo {year} {2006})}\BibitemShut {NoStop}%
\bibitem [{\citenamefont {Wei}\ \emph {et~al.}(2014)\citenamefont {Wei}, \citenamefont {Obstbaum}, \citenamefont {Ribow}, \citenamefont {Back},\ and\ \citenamefont {Woltersdorf}}]{Wei2014}%
  \BibitemOpen
  \bibfield  {author} {\bibinfo {author} {\bibfnamefont {D.}~\bibnamefont {Wei}}, \bibinfo {author} {\bibfnamefont {M.}~\bibnamefont {Obstbaum}}, \bibinfo {author} {\bibfnamefont {M.}~\bibnamefont {Ribow}}, \bibinfo {author} {\bibfnamefont {C.~H.}\ \bibnamefont {Back}},\ and\ \bibinfo {author} {\bibfnamefont {G.}~\bibnamefont {Woltersdorf}},\ }\bibfield  {title} {\bibinfo {title} {Spin hall voltages from a.c. and d.c. spin currents},\ }\href {https://doi.org/10.1038/ncomms4768} {\bibfield  {journal} {\bibinfo  {journal} {Nature Communications}\ }\textbf {\bibinfo {volume} {5}},\ \bibinfo {pages} {3768} (\bibinfo {year} {2014})}\BibitemShut {NoStop}%
\bibitem [{\citenamefont {Dushenko}\ \emph {et~al.}(2016)\citenamefont {Dushenko}, \citenamefont {Ago}, \citenamefont {Kawahara}, \citenamefont {Tsuda}, \citenamefont {Kuwabata}, \citenamefont {Takenobu}, \citenamefont {Shinjo}, \citenamefont {Ando},\ and\ \citenamefont {Shiraishi}}]{Dushenko2016}%
  \BibitemOpen
  \bibfield  {author} {\bibinfo {author} {\bibfnamefont {S.}~\bibnamefont {Dushenko}}, \bibinfo {author} {\bibfnamefont {H.}~\bibnamefont {Ago}}, \bibinfo {author} {\bibfnamefont {K.}~\bibnamefont {Kawahara}}, \bibinfo {author} {\bibfnamefont {T.}~\bibnamefont {Tsuda}}, \bibinfo {author} {\bibfnamefont {S.}~\bibnamefont {Kuwabata}}, \bibinfo {author} {\bibfnamefont {T.}~\bibnamefont {Takenobu}}, \bibinfo {author} {\bibfnamefont {T.}~\bibnamefont {Shinjo}}, \bibinfo {author} {\bibfnamefont {Y.}~\bibnamefont {Ando}},\ and\ \bibinfo {author} {\bibfnamefont {M.}~\bibnamefont {Shiraishi}},\ }\bibfield  {title} {\bibinfo {title} {Gate-tunable spin-charge conversion and the role of spin-orbit interaction in graphene},\ }\href {https://doi.org/10.1103/PhysRevLett.116.166102} {\bibfield  {journal} {\bibinfo  {journal} {Phys. Rev. Lett.}\ }\textbf {\bibinfo {volume} {116}},\ \bibinfo {pages} {166102} (\bibinfo {year} {2016})}\BibitemShut {NoStop}%
\bibitem [{\citenamefont {Lesne}\ \emph {et~al.}(2016)\citenamefont {Lesne}, \citenamefont {Fu}, \citenamefont {Oyarzun}, \citenamefont {Rojas-S{\'a}nchez}, \citenamefont {Vaz}, \citenamefont {Naganuma}, \citenamefont {Sicoli}, \citenamefont {Attan{\'e}}, \citenamefont {Jamet}, \citenamefont {Jacquet}, \citenamefont {George}, \citenamefont {Barth{\'e}l{\'e}my}, \citenamefont {Jaffr{\`e}s}, \citenamefont {Fert}, \citenamefont {Bibes},\ and\ \citenamefont {Vila}}]{Lesne2016}%
  \BibitemOpen
  \bibfield  {author} {\bibinfo {author} {\bibfnamefont {E.}~\bibnamefont {Lesne}}, \bibinfo {author} {\bibfnamefont {Y.}~\bibnamefont {Fu}}, \bibinfo {author} {\bibfnamefont {S.}~\bibnamefont {Oyarzun}}, \bibinfo {author} {\bibfnamefont {J.~C.}\ \bibnamefont {Rojas-S{\'a}nchez}}, \bibinfo {author} {\bibfnamefont {D.~C.}\ \bibnamefont {Vaz}}, \bibinfo {author} {\bibfnamefont {H.}~\bibnamefont {Naganuma}}, \bibinfo {author} {\bibfnamefont {G.}~\bibnamefont {Sicoli}}, \bibinfo {author} {\bibfnamefont {J.-P.}\ \bibnamefont {Attan{\'e}}}, \bibinfo {author} {\bibfnamefont {M.}~\bibnamefont {Jamet}}, \bibinfo {author} {\bibfnamefont {E.}~\bibnamefont {Jacquet}}, \bibinfo {author} {\bibfnamefont {J.-M.}\ \bibnamefont {George}}, \bibinfo {author} {\bibfnamefont {A.}~\bibnamefont {Barth{\'e}l{\'e}my}}, \bibinfo {author} {\bibfnamefont {H.}~\bibnamefont {Jaffr{\`e}s}}, \bibinfo {author} {\bibfnamefont {A.}~\bibnamefont {Fert}}, \bibinfo {author} {\bibfnamefont {M.}~\bibnamefont {Bibes}},\ and\ \bibinfo {author}
  {\bibfnamefont {L.}~\bibnamefont {Vila}},\ }\bibfield  {title} {\bibinfo {title} {Highly efficient and tunable spin-to-charge conversion through rashba coupling at oxide interfaces},\ }\href {https://doi.org/10.1038/nmat4726} {\bibfield  {journal} {\bibinfo  {journal} {Nature Materials}\ }\textbf {\bibinfo {volume} {15}},\ \bibinfo {pages} {1261} (\bibinfo {year} {2016})}\BibitemShut {NoStop}%
\bibitem [{\citenamefont {Kondou}\ \emph {et~al.}(2016)\citenamefont {Kondou}, \citenamefont {Yoshimi}, \citenamefont {Tsukazaki}, \citenamefont {Fukuma}, \citenamefont {Matsuno}, \citenamefont {Takahashi}, \citenamefont {Kawasaki}, \citenamefont {Tokura},\ and\ \citenamefont {Otani}}]{Kondou2016}%
  \BibitemOpen
  \bibfield  {author} {\bibinfo {author} {\bibfnamefont {K.}~\bibnamefont {Kondou}}, \bibinfo {author} {\bibfnamefont {R.}~\bibnamefont {Yoshimi}}, \bibinfo {author} {\bibfnamefont {A.}~\bibnamefont {Tsukazaki}}, \bibinfo {author} {\bibfnamefont {Y.}~\bibnamefont {Fukuma}}, \bibinfo {author} {\bibfnamefont {J.}~\bibnamefont {Matsuno}}, \bibinfo {author} {\bibfnamefont {K.~S.}\ \bibnamefont {Takahashi}}, \bibinfo {author} {\bibfnamefont {M.}~\bibnamefont {Kawasaki}}, \bibinfo {author} {\bibfnamefont {Y.}~\bibnamefont {Tokura}},\ and\ \bibinfo {author} {\bibfnamefont {Y.}~\bibnamefont {Otani}},\ }\bibfield  {title} {\bibinfo {title} {Fermi-level-dependent charge-to-spin current conversion by dirac surface states of topological insulators},\ }\href {https://doi.org/10.1038/nphys3833} {\bibfield  {journal} {\bibinfo  {journal} {Nature Physics}\ }\textbf {\bibinfo {volume} {12}},\ \bibinfo {pages} {1027} (\bibinfo {year} {2016})}\BibitemShut {NoStop}%
\bibitem [{\citenamefont {Hirsch}(1999)}]{Hirsch99}%
  \BibitemOpen
  \bibfield  {author} {\bibinfo {author} {\bibfnamefont {J.~E.}\ \bibnamefont {Hirsch}},\ }\bibfield  {title} {\bibinfo {title} {Spin hall effect},\ }\href {https://doi.org/10.1103/PhysRevLett.83.1834} {\bibfield  {journal} {\bibinfo  {journal} {Phys. Rev. Lett.}\ }\textbf {\bibinfo {volume} {83}},\ \bibinfo {pages} {1834} (\bibinfo {year} {1999})}\BibitemShut {NoStop}%
\bibitem [{\citenamefont {Zhang}(2000)}]{Zhang00}%
  \BibitemOpen
  \bibfield  {author} {\bibinfo {author} {\bibfnamefont {S.}~\bibnamefont {Zhang}},\ }\bibfield  {title} {\bibinfo {title} {Spin hall effect in the presence of spin diffusion},\ }\href {https://doi.org/10.1103/PhysRevLett.85.393} {\bibfield  {journal} {\bibinfo  {journal} {Phys. Rev. Lett.}\ }\textbf {\bibinfo {volume} {85}},\ \bibinfo {pages} {393} (\bibinfo {year} {2000})}\BibitemShut {NoStop}%
\bibitem [{\citenamefont {Murakami}\ \emph {et~al.}(2003)\citenamefont {Murakami}, \citenamefont {Nagaosa},\ and\ \citenamefont {Zhang}}]{Murakami2003}%
  \BibitemOpen
  \bibfield  {author} {\bibinfo {author} {\bibfnamefont {S.}~\bibnamefont {Murakami}}, \bibinfo {author} {\bibfnamefont {N.}~\bibnamefont {Nagaosa}},\ and\ \bibinfo {author} {\bibfnamefont {S.-C.}\ \bibnamefont {Zhang}},\ }\bibfield  {title} {\bibinfo {title} {Dissipationless quantum spin current at room temperature},\ }\href {https://doi.org/10.1126/science.1087128} {\bibfield  {journal} {\bibinfo  {journal} {Science}\ }\textbf {\bibinfo {volume} {301}},\ \bibinfo {pages} {1348} (\bibinfo {year} {2003})}\BibitemShut {NoStop}%
\bibitem [{\citenamefont {Sinova}\ \emph {et~al.}(2004)\citenamefont {Sinova}, \citenamefont {Culcer}, \citenamefont {Niu}, \citenamefont {Sinitsyn}, \citenamefont {Jungwirth},\ and\ \citenamefont {MacDonald}}]{Sinova04}%
  \BibitemOpen
  \bibfield  {author} {\bibinfo {author} {\bibfnamefont {J.}~\bibnamefont {Sinova}}, \bibinfo {author} {\bibfnamefont {D.}~\bibnamefont {Culcer}}, \bibinfo {author} {\bibfnamefont {Q.}~\bibnamefont {Niu}}, \bibinfo {author} {\bibfnamefont {N.~A.}\ \bibnamefont {Sinitsyn}}, \bibinfo {author} {\bibfnamefont {T.}~\bibnamefont {Jungwirth}},\ and\ \bibinfo {author} {\bibfnamefont {A.~H.}\ \bibnamefont {MacDonald}},\ }\bibfield  {title} {\bibinfo {title} {Universal intrinsic spin hall effect},\ }\href {https://doi.org/10.1103/PhysRevLett.92.126603} {\bibfield  {journal} {\bibinfo  {journal} {Phys. Rev. Lett.}\ }\textbf {\bibinfo {volume} {92}},\ \bibinfo {pages} {126603} (\bibinfo {year} {2004})}\BibitemShut {NoStop}%
\bibitem [{\citenamefont {Kato}\ \emph {et~al.}(2004)\citenamefont {Kato}, \citenamefont {Myers}, \citenamefont {Gossard},\ and\ \citenamefont {Awschalom}}]{Kato2004}%
  \BibitemOpen
  \bibfield  {author} {\bibinfo {author} {\bibfnamefont {Y.~K.}\ \bibnamefont {Kato}}, \bibinfo {author} {\bibfnamefont {R.~C.}\ \bibnamefont {Myers}}, \bibinfo {author} {\bibfnamefont {A.~C.}\ \bibnamefont {Gossard}},\ and\ \bibinfo {author} {\bibfnamefont {D.~D.}\ \bibnamefont {Awschalom}},\ }\bibfield  {title} {\bibinfo {title} {Observation of the spin hall effect in semiconductors},\ }\href {https://doi.org/10.1126/science.1105514} {\bibfield  {journal} {\bibinfo  {journal} {Science}\ }\textbf {\bibinfo {volume} {306}},\ \bibinfo {pages} {1910} (\bibinfo {year} {2004})}\BibitemShut {NoStop}%
\bibitem [{\citenamefont {Wunderlich}\ \emph {et~al.}(2005)\citenamefont {Wunderlich}, \citenamefont {Kaestner}, \citenamefont {Sinova},\ and\ \citenamefont {Jungwirth}}]{Wunderlich05}%
  \BibitemOpen
  \bibfield  {author} {\bibinfo {author} {\bibfnamefont {J.}~\bibnamefont {Wunderlich}}, \bibinfo {author} {\bibfnamefont {B.}~\bibnamefont {Kaestner}}, \bibinfo {author} {\bibfnamefont {J.}~\bibnamefont {Sinova}},\ and\ \bibinfo {author} {\bibfnamefont {T.}~\bibnamefont {Jungwirth}},\ }\bibfield  {title} {\bibinfo {title} {Experimental observation of the spin-hall effect in a two-dimensional spin-orbit coupled semiconductor system},\ }\href {https://doi.org/10.1103/PhysRevLett.94.047204} {\bibfield  {journal} {\bibinfo  {journal} {Phys. Rev. Lett.}\ }\textbf {\bibinfo {volume} {94}},\ \bibinfo {pages} {047204} (\bibinfo {year} {2005})}\BibitemShut {NoStop}%
\bibitem [{\citenamefont {Valenzuela}\ and\ \citenamefont {Tinkham}(2006)}]{Valenzuela2006}%
  \BibitemOpen
  \bibfield  {author} {\bibinfo {author} {\bibfnamefont {S.~O.}\ \bibnamefont {Valenzuela}}\ and\ \bibinfo {author} {\bibfnamefont {M.}~\bibnamefont {Tinkham}},\ }\bibfield  {title} {\bibinfo {title} {Direct electronic measurement of the spin hall effect},\ }\href {https://doi.org/10.1038/nature04937} {\bibfield  {journal} {\bibinfo  {journal} {Nature}\ }\textbf {\bibinfo {volume} {442}},\ \bibinfo {pages} {176} (\bibinfo {year} {2006})}\BibitemShut {NoStop}%
\bibitem [{\citenamefont {Yu}\ \emph {et~al.}(2014)\citenamefont {Yu}, \citenamefont {Wu}, \citenamefont {Liu}, \citenamefont {Xu},\ and\ \citenamefont {Yao}}]{Yu2014}%
  \BibitemOpen
  \bibfield  {author} {\bibinfo {author} {\bibfnamefont {H.}~\bibnamefont {Yu}}, \bibinfo {author} {\bibfnamefont {Y.}~\bibnamefont {Wu}}, \bibinfo {author} {\bibfnamefont {G.-B.}\ \bibnamefont {Liu}}, \bibinfo {author} {\bibfnamefont {X.}~\bibnamefont {Xu}},\ and\ \bibinfo {author} {\bibfnamefont {W.}~\bibnamefont {Yao}},\ }\bibfield  {title} {\bibinfo {title} {Nonlinear valley and spin currents from fermi pocket anisotropy in 2d crystals},\ }\href {https://doi.org/10.1103/PhysRevLett.113.156603} {\bibfield  {journal} {\bibinfo  {journal} {Phys. Rev. Lett.}\ }\textbf {\bibinfo {volume} {113}},\ \bibinfo {pages} {156603} (\bibinfo {year} {2014})}\BibitemShut {NoStop}%
\bibitem [{\citenamefont {Hamamoto}\ \emph {et~al.}(2017)\citenamefont {Hamamoto}, \citenamefont {Ezawa}, \citenamefont {Kim}, \citenamefont {Morimoto},\ and\ \citenamefont {Nagaosa}}]{Hamamoto2017}%
  \BibitemOpen
  \bibfield  {author} {\bibinfo {author} {\bibfnamefont {K.}~\bibnamefont {Hamamoto}}, \bibinfo {author} {\bibfnamefont {M.}~\bibnamefont {Ezawa}}, \bibinfo {author} {\bibfnamefont {K.~W.}\ \bibnamefont {Kim}}, \bibinfo {author} {\bibfnamefont {T.}~\bibnamefont {Morimoto}},\ and\ \bibinfo {author} {\bibfnamefont {N.}~\bibnamefont {Nagaosa}},\ }\bibfield  {title} {\bibinfo {title} {Nonlinear spin current generation in noncentrosymmetric spin-orbit coupled systems},\ }\href {https://doi.org/10.1103/PhysRevB.95.224430} {\bibfield  {journal} {\bibinfo  {journal} {Phys. Rev. B}\ }\textbf {\bibinfo {volume} {95}},\ \bibinfo {pages} {224430} (\bibinfo {year} {2017})}\BibitemShut {NoStop}%
\bibitem [{\citenamefont {Barman}\ \emph {et~al.}(2021)\citenamefont {Barman}, \citenamefont {Gubbiotti}, \citenamefont {Ladak}, \citenamefont {Adeyeye}, \citenamefont {Krawczyk}, \citenamefont {Gr{\"a}fe}, \citenamefont {Adelmann}, \citenamefont {Cotofana}, \citenamefont {Naeemi}, \citenamefont {Vasyuchka}, \citenamefont {Hillebrands}, \citenamefont {Nikitov}, \citenamefont {Yu}, \citenamefont {Grundler}, \citenamefont {Sadovnikov}, \citenamefont {Grachev}, \citenamefont {Sheshukova}, \citenamefont {Duquesne}, \citenamefont {Marangolo}, \citenamefont {Csaba}, \citenamefont {Porod}, \citenamefont {Demidov}, \citenamefont {Urazhdin}, \citenamefont {Demokritov}, \citenamefont {Albisetti}, \citenamefont {Petti}, \citenamefont {Bertacco}, \citenamefont {Schultheiss}, \citenamefont {Kruglyak}, \citenamefont {Poimanov}, \citenamefont {Sahoo}, \citenamefont {Sinha}, \citenamefont {Yang}, \citenamefont {M{\"u}nzenberg}, \citenamefont {Moriyama}, \citenamefont {Mizukami}, \citenamefont {Landeros}, \citenamefont
  {Gallardo}, \citenamefont {Carlotti}, \citenamefont {Kim}, \citenamefont {Stamps}, \citenamefont {Camley}, \citenamefont {Rana}, \citenamefont {Otani}, \citenamefont {Yu}, \citenamefont {Yu}, \citenamefont {Bauer}, \citenamefont {Back}, \citenamefont {Uhrig}, \citenamefont {Dobrovolskiy}, \citenamefont {Budinska}, \citenamefont {Qin}, \citenamefont {van Dijken}, \citenamefont {Chumak}, \citenamefont {Khitun}, \citenamefont {Nikonov}, \citenamefont {Young}, \citenamefont {Zingsem},\ and\ \citenamefont {Winklhofer}}]{Barman2021}%
  \BibitemOpen
  \bibfield  {author} {\bibinfo {author} {\bibfnamefont {A.}~\bibnamefont {Barman}}, \bibinfo {author} {\bibfnamefont {G.}~\bibnamefont {Gubbiotti}}, \bibinfo {author} {\bibfnamefont {S.}~\bibnamefont {Ladak}}, \bibinfo {author} {\bibfnamefont {A.~O.}\ \bibnamefont {Adeyeye}}, \bibinfo {author} {\bibfnamefont {M.}~\bibnamefont {Krawczyk}}, \bibinfo {author} {\bibfnamefont {J.}~\bibnamefont {Gr{\"a}fe}}, \bibinfo {author} {\bibfnamefont {C.}~\bibnamefont {Adelmann}}, \bibinfo {author} {\bibfnamefont {S.}~\bibnamefont {Cotofana}}, \bibinfo {author} {\bibfnamefont {A.}~\bibnamefont {Naeemi}}, \bibinfo {author} {\bibfnamefont {V.~I.}\ \bibnamefont {Vasyuchka}}, \bibinfo {author} {\bibfnamefont {B.}~\bibnamefont {Hillebrands}}, \bibinfo {author} {\bibfnamefont {S.~A.}\ \bibnamefont {Nikitov}}, \bibinfo {author} {\bibfnamefont {H.}~\bibnamefont {Yu}}, \bibinfo {author} {\bibfnamefont {D.}~\bibnamefont {Grundler}}, \bibinfo {author} {\bibfnamefont {A.~V.}\ \bibnamefont {Sadovnikov}}, \bibinfo {author} {\bibfnamefont
  {A.~A.}\ \bibnamefont {Grachev}}, \bibinfo {author} {\bibfnamefont {S.~E.}\ \bibnamefont {Sheshukova}}, \bibinfo {author} {\bibfnamefont {J.-Y.}\ \bibnamefont {Duquesne}}, \bibinfo {author} {\bibfnamefont {M.}~\bibnamefont {Marangolo}}, \bibinfo {author} {\bibfnamefont {G.}~\bibnamefont {Csaba}}, \bibinfo {author} {\bibfnamefont {W.}~\bibnamefont {Porod}}, \bibinfo {author} {\bibfnamefont {V.~E.}\ \bibnamefont {Demidov}}, \bibinfo {author} {\bibfnamefont {S.}~\bibnamefont {Urazhdin}}, \bibinfo {author} {\bibfnamefont {S.~O.}\ \bibnamefont {Demokritov}}, \bibinfo {author} {\bibfnamefont {E.}~\bibnamefont {Albisetti}}, \bibinfo {author} {\bibfnamefont {D.}~\bibnamefont {Petti}}, \bibinfo {author} {\bibfnamefont {R.}~\bibnamefont {Bertacco}}, \bibinfo {author} {\bibfnamefont {H.}~\bibnamefont {Schultheiss}}, \bibinfo {author} {\bibfnamefont {V.~V.}\ \bibnamefont {Kruglyak}}, \bibinfo {author} {\bibfnamefont {V.~D.}\ \bibnamefont {Poimanov}}, \bibinfo {author} {\bibfnamefont {S.}~\bibnamefont {Sahoo}}, \bibinfo
  {author} {\bibfnamefont {J.}~\bibnamefont {Sinha}}, \bibinfo {author} {\bibfnamefont {H.}~\bibnamefont {Yang}}, \bibinfo {author} {\bibfnamefont {M.}~\bibnamefont {M{\"u}nzenberg}}, \bibinfo {author} {\bibfnamefont {T.}~\bibnamefont {Moriyama}}, \bibinfo {author} {\bibfnamefont {S.}~\bibnamefont {Mizukami}}, \bibinfo {author} {\bibfnamefont {P.}~\bibnamefont {Landeros}}, \bibinfo {author} {\bibfnamefont {R.~A.}\ \bibnamefont {Gallardo}}, \bibinfo {author} {\bibfnamefont {G.}~\bibnamefont {Carlotti}}, \bibinfo {author} {\bibfnamefont {J.-V.}\ \bibnamefont {Kim}}, \bibinfo {author} {\bibfnamefont {R.~L.}\ \bibnamefont {Stamps}}, \bibinfo {author} {\bibfnamefont {R.~E.}\ \bibnamefont {Camley}}, \bibinfo {author} {\bibfnamefont {B.}~\bibnamefont {Rana}}, \bibinfo {author} {\bibfnamefont {Y.}~\bibnamefont {Otani}}, \bibinfo {author} {\bibfnamefont {W.}~\bibnamefont {Yu}}, \bibinfo {author} {\bibfnamefont {T.}~\bibnamefont {Yu}}, \bibinfo {author} {\bibfnamefont {G.~E.~W.}\ \bibnamefont {Bauer}}, \bibinfo
  {author} {\bibfnamefont {C.}~\bibnamefont {Back}}, \bibinfo {author} {\bibfnamefont {G.~S.}\ \bibnamefont {Uhrig}}, \bibinfo {author} {\bibfnamefont {O.~V.}\ \bibnamefont {Dobrovolskiy}}, \bibinfo {author} {\bibfnamefont {B.}~\bibnamefont {Budinska}}, \bibinfo {author} {\bibfnamefont {H.}~\bibnamefont {Qin}}, \bibinfo {author} {\bibfnamefont {S.}~\bibnamefont {van Dijken}}, \bibinfo {author} {\bibfnamefont {A.~V.}\ \bibnamefont {Chumak}}, \bibinfo {author} {\bibfnamefont {A.}~\bibnamefont {Khitun}}, \bibinfo {author} {\bibfnamefont {D.~E.}\ \bibnamefont {Nikonov}}, \bibinfo {author} {\bibfnamefont {I.~A.}\ \bibnamefont {Young}}, \bibinfo {author} {\bibfnamefont {B.~W.}\ \bibnamefont {Zingsem}},\ and\ \bibinfo {author} {\bibfnamefont {M.}~\bibnamefont {Winklhofer}},\ }\bibfield  {title} {\bibinfo {title} {The 2021 magnonics roadmap},\ }\href {https://doi.org/10.1088/1361-648X/abec1a} {\bibfield  {journal} {\bibinfo  {journal} {Journal of Physics: Condensed Matter}\ }\textbf {\bibinfo {volume} {33}},\
  \bibinfo {pages} {413001} (\bibinfo {year} {2021})}\BibitemShut {NoStop}%
\bibitem [{\citenamefont {Demokritov}\ and\ \citenamefont {Slavin}(2013)}]{Democritov2013}%
  \BibitemOpen
  \bibfield  {author} {\bibinfo {author} {\bibfnamefont {S.~O.}\ \bibnamefont {Demokritov}}\ and\ \bibinfo {author} {\bibfnamefont {A.~N.}\ \bibnamefont {Slavin}},\ }\href@noop {} {\emph {\bibinfo {title} {Magnonics from Fundamentals to Applications}}}\ (\bibinfo  {publisher} {Berlin: Springer},\ \bibinfo {year} {2013})\BibitemShut {NoStop}%
\bibitem [{\citenamefont {Rezende}(2020)}]{Rezende2020}%
  \BibitemOpen
  \bibfield  {author} {\bibinfo {author} {\bibfnamefont {S.}~\bibnamefont {Rezende}},\ }\href@noop {} {\emph {\bibinfo {title} {Fundamentals of Magnonics}}}\ (\bibinfo  {publisher} {Berlin: Springer},\ \bibinfo {year} {2020})\BibitemShut {NoStop}%
\bibitem [{\citenamefont {Brataas}\ \emph {et~al.}(2020)\citenamefont {Brataas}, \citenamefont {van Wees}, \citenamefont {Klein}, \citenamefont {de~Loubens},\ and\ \citenamefont {Viret}}]{Brataas2020}%
  \BibitemOpen
  \bibfield  {author} {\bibinfo {author} {\bibfnamefont {A.}~\bibnamefont {Brataas}}, \bibinfo {author} {\bibfnamefont {B.}~\bibnamefont {van Wees}}, \bibinfo {author} {\bibfnamefont {O.}~\bibnamefont {Klein}}, \bibinfo {author} {\bibfnamefont {G.}~\bibnamefont {de~Loubens}},\ and\ \bibinfo {author} {\bibfnamefont {M.}~\bibnamefont {Viret}},\ }\bibfield  {title} {\bibinfo {title} {Spin insulatronics},\ }\href {https://www.sciencedirect.com/science/article/pii/S0370157320302933} {\bibfield  {journal} {\bibinfo  {journal} {Physics Reports}\ }\textbf {\bibinfo {volume} {885}},\ \bibinfo {pages} {1} (\bibinfo {year} {2020})}\BibitemShut {NoStop}%
\bibitem [{\citenamefont {Chumak}\ \emph {et~al.}(2015)\citenamefont {Chumak}, \citenamefont {Vasyuchka}, \citenamefont {Serga},\ and\ \citenamefont {Hillebrands}}]{Chumak2015}%
  \BibitemOpen
  \bibfield  {author} {\bibinfo {author} {\bibfnamefont {A.~V.}\ \bibnamefont {Chumak}}, \bibinfo {author} {\bibfnamefont {V.~I.}\ \bibnamefont {Vasyuchka}}, \bibinfo {author} {\bibfnamefont {A.~A.}\ \bibnamefont {Serga}},\ and\ \bibinfo {author} {\bibfnamefont {B.}~\bibnamefont {Hillebrands}},\ }\bibfield  {title} {\bibinfo {title} {Magnon spintronics},\ }\href {https://doi.org/10.1038/nphys3347} {\bibfield  {journal} {\bibinfo  {journal} {Nature Physics}\ }\textbf {\bibinfo {volume} {11}},\ \bibinfo {pages} {453} (\bibinfo {year} {2015})}\BibitemShut {NoStop}%
\bibitem [{\citenamefont {Linder}\ and\ \citenamefont {Robinson}(2015)}]{Linder2015}%
  \BibitemOpen
  \bibfield  {author} {\bibinfo {author} {\bibfnamefont {J.}~\bibnamefont {Linder}}\ and\ \bibinfo {author} {\bibfnamefont {J.~W.~A.}\ \bibnamefont {Robinson}},\ }\bibfield  {title} {\bibinfo {title} {Superconducting spintronics},\ }\href {https://doi.org/10.1038/nphys3242} {\bibfield  {journal} {\bibinfo  {journal} {Nature Physics}\ }\textbf {\bibinfo {volume} {11}},\ \bibinfo {pages} {307} (\bibinfo {year} {2015})}\BibitemShut {NoStop}%
\bibitem [{\citenamefont {Eschrig}(2015)}]{Eschrig2015}%
  \BibitemOpen
  \bibfield  {author} {\bibinfo {author} {\bibfnamefont {M.}~\bibnamefont {Eschrig}},\ }\bibfield  {title} {\bibinfo {title} {Spin-polarized supercurrents for spintronics: a review of current progress},\ }\href {https://doi.org/10.1088/0034-4885/78/10/104501} {\bibfield  {journal} {\bibinfo  {journal} {Reports on Progress in Physics}\ }\textbf {\bibinfo {volume} {78}},\ \bibinfo {pages} {104501} (\bibinfo {year} {2015})}\BibitemShut {NoStop}%
\bibitem [{\citenamefont {Leggett}(1975)}]{Leggett1975}%
  \BibitemOpen
  \bibfield  {author} {\bibinfo {author} {\bibfnamefont {A.~J.}\ \bibnamefont {Leggett}},\ }\bibfield  {title} {\bibinfo {title} {A theoretical description of the new phases of liquid $^{3}\mathrm{He}$},\ }\href {https://doi.org/10.1103/RevModPhys.47.331} {\bibfield  {journal} {\bibinfo  {journal} {Rev. Mod. Phys.}\ }\textbf {\bibinfo {volume} {47}},\ \bibinfo {pages} {331} (\bibinfo {year} {1975})}\BibitemShut {NoStop}%
\bibitem [{\citenamefont {Asano}(2005)}]{Asano2005}%
  \BibitemOpen
  \bibfield  {author} {\bibinfo {author} {\bibfnamefont {Y.}~\bibnamefont {Asano}},\ }\bibfield  {title} {\bibinfo {title} {Spin current in $p$-wave superconducting rings},\ }\href {https://doi.org/10.1103/PhysRevB.72.092508} {\bibfield  {journal} {\bibinfo  {journal} {Phys. Rev. B}\ }\textbf {\bibinfo {volume} {72}},\ \bibinfo {pages} {092508} (\bibinfo {year} {2005})}\BibitemShut {NoStop}%
\bibitem [{\citenamefont {Asano}(2006)}]{Asano2006}%
  \BibitemOpen
  \bibfield  {author} {\bibinfo {author} {\bibfnamefont {Y.}~\bibnamefont {Asano}},\ }\bibfield  {title} {\bibinfo {title} {Josephson spin current in triplet superconductor junctions},\ }\href {https://doi.org/10.1103/PhysRevB.74.220501} {\bibfield  {journal} {\bibinfo  {journal} {Phys. Rev. B}\ }\textbf {\bibinfo {volume} {74}},\ \bibinfo {pages} {220501} (\bibinfo {year} {2006})}\BibitemShut {NoStop}%
\bibitem [{\citenamefont {Leurs}\ \emph {et~al.}(2008)\citenamefont {Leurs}, \citenamefont {Nazario}, \citenamefont {Santiago},\ and\ \citenamefont {Zaanen}}]{Leurs2008}%
  \BibitemOpen
  \bibfield  {author} {\bibinfo {author} {\bibfnamefont {B.~W.~A.}\ \bibnamefont {Leurs}}, \bibinfo {author} {\bibfnamefont {Z.}~\bibnamefont {Nazario}}, \bibinfo {author} {\bibfnamefont {D.~I.}\ \bibnamefont {Santiago}},\ and\ \bibinfo {author} {\bibfnamefont {J.}~\bibnamefont {Zaanen}},\ }\bibfield  {title} {\bibinfo {title} {Non-abelian hydrodynamics and the flow of spin in spin--orbit coupled substances},\ }\href {https://www.sciencedirect.com/science/article/pii/S0003491607000863} {\bibfield  {journal} {\bibinfo  {journal} {Annals of Physics}\ }\textbf {\bibinfo {volume} {323}},\ \bibinfo {pages} {907} (\bibinfo {year} {2008})}\BibitemShut {NoStop}%
\bibitem [{\citenamefont {He}\ \emph {et~al.}(2019)\citenamefont {He}, \citenamefont {Hiroki}, \citenamefont {Hamamoto},\ and\ \citenamefont {Nagaosa}}]{He2019}%
  \BibitemOpen
  \bibfield  {author} {\bibinfo {author} {\bibfnamefont {J.~J.}\ \bibnamefont {He}}, \bibinfo {author} {\bibfnamefont {K.}~\bibnamefont {Hiroki}}, \bibinfo {author} {\bibfnamefont {K.}~\bibnamefont {Hamamoto}},\ and\ \bibinfo {author} {\bibfnamefont {N.}~\bibnamefont {Nagaosa}},\ }\bibfield  {title} {\bibinfo {title} {Spin supercurrent in two-dimensional superconductors with rashba spin-orbit interaction},\ }\href {https://doi.org/10.1038/s42005-019-0230-9} {\bibfield  {journal} {\bibinfo  {journal} {Communications Physics}\ }\textbf {\bibinfo {volume} {2}},\ \bibinfo {pages} {128} (\bibinfo {year} {2019})}\BibitemShut {NoStop}%
\bibitem [{\citenamefont {Bergeret}\ and\ \citenamefont {Tokatly}(2016)}]{Bergeret2016}%
  \BibitemOpen
  \bibfield  {author} {\bibinfo {author} {\bibfnamefont {F.~S.}\ \bibnamefont {Bergeret}}\ and\ \bibinfo {author} {\bibfnamefont {I.~V.}\ \bibnamefont {Tokatly}},\ }\bibfield  {title} {\bibinfo {title} {Manifestation of extrinsic spin hall effect in superconducting structures: Nondissipative magnetoelectric effects},\ }\href {https://doi.org/10.1103/PhysRevB.94.180502} {\bibfield  {journal} {\bibinfo  {journal} {Phys. Rev. B}\ }\textbf {\bibinfo {volume} {94}},\ \bibinfo {pages} {180502} (\bibinfo {year} {2016})}\BibitemShut {NoStop}%
\bibitem [{\citenamefont {Grein}\ \emph {et~al.}(2009)\citenamefont {Grein}, \citenamefont {Eschrig}, \citenamefont {Metalidis},\ and\ \citenamefont {Sch\"on}}]{Grein2009}%
  \BibitemOpen
  \bibfield  {author} {\bibinfo {author} {\bibfnamefont {R.}~\bibnamefont {Grein}}, \bibinfo {author} {\bibfnamefont {M.}~\bibnamefont {Eschrig}}, \bibinfo {author} {\bibfnamefont {G.}~\bibnamefont {Metalidis}},\ and\ \bibinfo {author} {\bibfnamefont {G.}~\bibnamefont {Sch\"on}},\ }\bibfield  {title} {\bibinfo {title} {Spin-dependent cooper pair phase and pure spin supercurrents in strongly polarized ferromagnets},\ }\href {https://doi.org/10.1103/PhysRevLett.102.227005} {\bibfield  {journal} {\bibinfo  {journal} {Phys. Rev. Lett.}\ }\textbf {\bibinfo {volume} {102}},\ \bibinfo {pages} {227005} (\bibinfo {year} {2009})}\BibitemShut {NoStop}%
\bibitem [{\citenamefont {Jeon}\ \emph {et~al.}(2020)\citenamefont {Jeon}, \citenamefont {Montiel}, \citenamefont {Komori}, \citenamefont {Ciccarelli}, \citenamefont {Haigh}, \citenamefont {Kurebayashi}, \citenamefont {Cohen}, \citenamefont {Chan}, \citenamefont {Stenning}, \citenamefont {Lee}, \citenamefont {Eschrig}, \citenamefont {Blamire},\ and\ \citenamefont {Robinson}}]{Jeon2020}%
  \BibitemOpen
  \bibfield  {author} {\bibinfo {author} {\bibfnamefont {K.-R.}\ \bibnamefont {Jeon}}, \bibinfo {author} {\bibfnamefont {X.}~\bibnamefont {Montiel}}, \bibinfo {author} {\bibfnamefont {S.}~\bibnamefont {Komori}}, \bibinfo {author} {\bibfnamefont {C.}~\bibnamefont {Ciccarelli}}, \bibinfo {author} {\bibfnamefont {J.}~\bibnamefont {Haigh}}, \bibinfo {author} {\bibfnamefont {H.}~\bibnamefont {Kurebayashi}}, \bibinfo {author} {\bibfnamefont {L.~F.}\ \bibnamefont {Cohen}}, \bibinfo {author} {\bibfnamefont {A.~K.}\ \bibnamefont {Chan}}, \bibinfo {author} {\bibfnamefont {K.~D.}\ \bibnamefont {Stenning}}, \bibinfo {author} {\bibfnamefont {C.-M.}\ \bibnamefont {Lee}}, \bibinfo {author} {\bibfnamefont {M.}~\bibnamefont {Eschrig}}, \bibinfo {author} {\bibfnamefont {M.~G.}\ \bibnamefont {Blamire}},\ and\ \bibinfo {author} {\bibfnamefont {J.~W.~A.}\ \bibnamefont {Robinson}},\ }\bibfield  {title} {\bibinfo {title} {Tunable pure spin supercurrents and the demonstration of their gateability in a spin-wave device},\ }\href
  {https://doi.org/10.1103/PhysRevX.10.031020} {\bibfield  {journal} {\bibinfo  {journal} {Phys. Rev. X}\ }\textbf {\bibinfo {volume} {10}},\ \bibinfo {pages} {031020} (\bibinfo {year} {2020})}\BibitemShut {NoStop}%
\bibitem [{\citenamefont {Brydon}\ \emph {et~al.}(2013)\citenamefont {Brydon}, \citenamefont {Chen}, \citenamefont {Asano},\ and\ \citenamefont {Manske}}]{Brydon2013}%
  \BibitemOpen
  \bibfield  {author} {\bibinfo {author} {\bibfnamefont {P.~M.~R.}\ \bibnamefont {Brydon}}, \bibinfo {author} {\bibfnamefont {W.}~\bibnamefont {Chen}}, \bibinfo {author} {\bibfnamefont {Y.}~\bibnamefont {Asano}},\ and\ \bibinfo {author} {\bibfnamefont {D.}~\bibnamefont {Manske}},\ }\bibfield  {title} {\bibinfo {title} {Charge and spin supercurrents in triplet superconductor-ferromagnet-singlet superconductor josephson junctions},\ }\href {https://doi.org/10.1103/PhysRevB.88.054509} {\bibfield  {journal} {\bibinfo  {journal} {Phys. Rev. B}\ }\textbf {\bibinfo {volume} {88}},\ \bibinfo {pages} {054509} (\bibinfo {year} {2013})}\BibitemShut {NoStop}%
\bibitem [{\citenamefont {Dahir}\ \emph {et~al.}(2022)\citenamefont {Dahir}, \citenamefont {Volkov},\ and\ \citenamefont {Eremin}}]{Dahir2022}%
  \BibitemOpen
  \bibfield  {author} {\bibinfo {author} {\bibfnamefont {S.~M.}\ \bibnamefont {Dahir}}, \bibinfo {author} {\bibfnamefont {A.~F.}\ \bibnamefont {Volkov}},\ and\ \bibinfo {author} {\bibfnamefont {I.~M.}\ \bibnamefont {Eremin}},\ }\bibfield  {title} {\bibinfo {title} {Charge and spin supercurrents in magnetic josephson junctions with spin filters and domain walls},\ }\href {https://doi.org/10.1103/PhysRevB.105.094517} {\bibfield  {journal} {\bibinfo  {journal} {Phys. Rev. B}\ }\textbf {\bibinfo {volume} {105}},\ \bibinfo {pages} {094517} (\bibinfo {year} {2022})}\BibitemShut {NoStop}%
\bibitem [{\citenamefont {Ojaj\"arvi}\ \emph {et~al.}(2022)\citenamefont {Ojaj\"arvi}, \citenamefont {Bergeret}, \citenamefont {Silaev},\ and\ \citenamefont {Heikkil\"a}}]{Ojajarvi2022}%
  \BibitemOpen
  \bibfield  {author} {\bibinfo {author} {\bibfnamefont {R.}~\bibnamefont {Ojaj\"arvi}}, \bibinfo {author} {\bibfnamefont {F.~S.}\ \bibnamefont {Bergeret}}, \bibinfo {author} {\bibfnamefont {M.~A.}\ \bibnamefont {Silaev}},\ and\ \bibinfo {author} {\bibfnamefont {T.~T.}\ \bibnamefont {Heikkil\"a}},\ }\bibfield  {title} {\bibinfo {title} {Dynamics of two ferromagnetic insulators coupled by superconducting spin current},\ }\href {https://doi.org/10.1103/PhysRevLett.128.167701} {\bibfield  {journal} {\bibinfo  {journal} {Phys. Rev. Lett.}\ }\textbf {\bibinfo {volume} {128}},\ \bibinfo {pages} {167701} (\bibinfo {year} {2022})}\BibitemShut {NoStop}%
\bibitem [{\citenamefont {Ouassou}\ \emph {et~al.}(2017)\citenamefont {Ouassou}, \citenamefont {Jacobsen},\ and\ \citenamefont {Linder}}]{Ouassou2017}%
  \BibitemOpen
  \bibfield  {author} {\bibinfo {author} {\bibfnamefont {J.~A.}\ \bibnamefont {Ouassou}}, \bibinfo {author} {\bibfnamefont {S.~H.}\ \bibnamefont {Jacobsen}},\ and\ \bibinfo {author} {\bibfnamefont {J.}~\bibnamefont {Linder}},\ }\bibfield  {title} {\bibinfo {title} {Conservation of spin supercurrents in superconductors},\ }\href {https://doi.org/10.1103/PhysRevB.96.094505} {\bibfield  {journal} {\bibinfo  {journal} {Phys. Rev. B}\ }\textbf {\bibinfo {volume} {96}},\ \bibinfo {pages} {094505} (\bibinfo {year} {2017})}\BibitemShut {NoStop}%
\bibitem [{\citenamefont {Ouassou}\ \emph {et~al.}(2019)\citenamefont {Ouassou}, \citenamefont {Robinson},\ and\ \citenamefont {Linder}}]{Ouassou2019}%
  \BibitemOpen
  \bibfield  {author} {\bibinfo {author} {\bibfnamefont {J.~A.}\ \bibnamefont {Ouassou}}, \bibinfo {author} {\bibfnamefont {J.~W.~A.}\ \bibnamefont {Robinson}},\ and\ \bibinfo {author} {\bibfnamefont {J.}~\bibnamefont {Linder}},\ }\bibfield  {title} {\bibinfo {title} {Controlling spin supercurrents via nonequilibrium spin injection},\ }\href {https://doi.org/10.1038/s41598-019-48945-0} {\bibfield  {journal} {\bibinfo  {journal} {Scientific Reports}\ }\textbf {\bibinfo {volume} {9}},\ \bibinfo {pages} {12731} (\bibinfo {year} {2019})}\BibitemShut {NoStop}%
\bibitem [{\citenamefont {Alidoust}\ \emph {et~al.}(2010)\citenamefont {Alidoust}, \citenamefont {Linder}, \citenamefont {Rashedi}, \citenamefont {Yokoyama},\ and\ \citenamefont {Sudb\o{}}}]{Alidoust2010}%
  \BibitemOpen
  \bibfield  {author} {\bibinfo {author} {\bibfnamefont {M.}~\bibnamefont {Alidoust}}, \bibinfo {author} {\bibfnamefont {J.}~\bibnamefont {Linder}}, \bibinfo {author} {\bibfnamefont {G.}~\bibnamefont {Rashedi}}, \bibinfo {author} {\bibfnamefont {T.}~\bibnamefont {Yokoyama}},\ and\ \bibinfo {author} {\bibfnamefont {A.}~\bibnamefont {Sudb\o{}}},\ }\bibfield  {title} {\bibinfo {title} {Spin-polarized josephson current in superconductor/ferromagnet/superconductor junctions with inhomogeneous magnetization},\ }\href {https://doi.org/10.1103/PhysRevB.81.014512} {\bibfield  {journal} {\bibinfo  {journal} {Phys. Rev. B}\ }\textbf {\bibinfo {volume} {81}},\ \bibinfo {pages} {014512} (\bibinfo {year} {2010})}\BibitemShut {NoStop}%
\bibitem [{\citenamefont {Jacobsen}\ \emph {et~al.}(2016)\citenamefont {Jacobsen}, \citenamefont {Kulagina},\ and\ \citenamefont {Linder}}]{Jacobsen2016}%
  \BibitemOpen
  \bibfield  {author} {\bibinfo {author} {\bibfnamefont {S.~H.}\ \bibnamefont {Jacobsen}}, \bibinfo {author} {\bibfnamefont {I.}~\bibnamefont {Kulagina}},\ and\ \bibinfo {author} {\bibfnamefont {J.}~\bibnamefont {Linder}},\ }\bibfield  {title} {\bibinfo {title} {Controlling superconducting spin flow with spin-flip immunity using a single homogeneous ferromagnet},\ }\href {https://doi.org/10.1038/srep23926} {\bibfield  {journal} {\bibinfo  {journal} {Scientific Reports}\ }\textbf {\bibinfo {volume} {6}},\ \bibinfo {pages} {23926} (\bibinfo {year} {2016})}\BibitemShut {NoStop}%
\bibitem [{\citenamefont {Gomperud}\ and\ \citenamefont {Linder}(2015)}]{Gomperud2015}%
  \BibitemOpen
  \bibfield  {author} {\bibinfo {author} {\bibfnamefont {I.}~\bibnamefont {Gomperud}}\ and\ \bibinfo {author} {\bibfnamefont {J.}~\bibnamefont {Linder}},\ }\bibfield  {title} {\bibinfo {title} {Spin supercurrent and phase-tunable triplet cooper pairs via magnetic insulators},\ }\href {https://doi.org/10.1103/PhysRevB.92.035416} {\bibfield  {journal} {\bibinfo  {journal} {Phys. Rev. B}\ }\textbf {\bibinfo {volume} {92}},\ \bibinfo {pages} {035416} (\bibinfo {year} {2015})}\BibitemShut {NoStop}%
\bibitem [{\citenamefont {Brydon}\ \emph {et~al.}(2011)\citenamefont {Brydon}, \citenamefont {Asano},\ and\ \citenamefont {Timm}}]{Brydon2011}%
  \BibitemOpen
  \bibfield  {author} {\bibinfo {author} {\bibfnamefont {P.~M.~R.}\ \bibnamefont {Brydon}}, \bibinfo {author} {\bibfnamefont {Y.}~\bibnamefont {Asano}},\ and\ \bibinfo {author} {\bibfnamefont {C.}~\bibnamefont {Timm}},\ }\bibfield  {title} {\bibinfo {title} {Spin josephson effect with a single superconductor},\ }\href {https://doi.org/10.1103/PhysRevB.83.180504} {\bibfield  {journal} {\bibinfo  {journal} {Phys. Rev. B}\ }\textbf {\bibinfo {volume} {83}},\ \bibinfo {pages} {180504} (\bibinfo {year} {2011})}\BibitemShut {NoStop}%
\bibitem [{\citenamefont {Linder}\ \emph {et~al.}(2017)\citenamefont {Linder}, \citenamefont {Amundsen},\ and\ \citenamefont {Risingg\aa{}rd}}]{Linder2017}%
  \BibitemOpen
  \bibfield  {author} {\bibinfo {author} {\bibfnamefont {J.}~\bibnamefont {Linder}}, \bibinfo {author} {\bibfnamefont {M.}~\bibnamefont {Amundsen}},\ and\ \bibinfo {author} {\bibfnamefont {V.}~\bibnamefont {Risingg\aa{}rd}},\ }\bibfield  {title} {\bibinfo {title} {Intrinsic superspin hall current},\ }\href {https://doi.org/10.1103/PhysRevB.96.094512} {\bibfield  {journal} {\bibinfo  {journal} {Phys. Rev. B}\ }\textbf {\bibinfo {volume} {96}},\ \bibinfo {pages} {094512} (\bibinfo {year} {2017})}\BibitemShut {NoStop}%
\bibitem [{\citenamefont {Bobkova}\ \emph {et~al.}(2017)\citenamefont {Bobkova}, \citenamefont {Bobkov},\ and\ \citenamefont {Silaev}}]{Bobkova2017}%
  \BibitemOpen
  \bibfield  {author} {\bibinfo {author} {\bibfnamefont {I.~V.}\ \bibnamefont {Bobkova}}, \bibinfo {author} {\bibfnamefont {A.~M.}\ \bibnamefont {Bobkov}},\ and\ \bibinfo {author} {\bibfnamefont {M.~A.}\ \bibnamefont {Silaev}},\ }\bibfield  {title} {\bibinfo {title} {Gauge theory of the long-range proximity effect and spontaneous currents in superconducting heterostructures with strong ferromagnets},\ }\href {https://doi.org/10.1103/PhysRevB.96.094506} {\bibfield  {journal} {\bibinfo  {journal} {Phys. Rev. B}\ }\textbf {\bibinfo {volume} {96}},\ \bibinfo {pages} {094506} (\bibinfo {year} {2017})}\BibitemShut {NoStop}%
\bibitem [{\citenamefont {Bobkova}\ \emph {et~al.}(2018)\citenamefont {Bobkova}, \citenamefont {Bobkov},\ and\ \citenamefont {Silaev}}]{Bobkova2018}%
  \BibitemOpen
  \bibfield  {author} {\bibinfo {author} {\bibfnamefont {I.~V.}\ \bibnamefont {Bobkova}}, \bibinfo {author} {\bibfnamefont {A.~M.}\ \bibnamefont {Bobkov}},\ and\ \bibinfo {author} {\bibfnamefont {M.~A.}\ \bibnamefont {Silaev}},\ }\bibfield  {title} {\bibinfo {title} {Spin torques and magnetic texture dynamics driven by the supercurrent in superconductor/ferromagnet structures},\ }\href {https://doi.org/10.1103/PhysRevB.98.014521} {\bibfield  {journal} {\bibinfo  {journal} {Phys. Rev. B}\ }\textbf {\bibinfo {volume} {98}},\ \bibinfo {pages} {014521} (\bibinfo {year} {2018})}\BibitemShut {NoStop}%
\bibitem [{\citenamefont {Aunsmo}\ and\ \citenamefont {Linder}(2024)}]{Aunsmo2024}%
  \BibitemOpen
  \bibfield  {author} {\bibinfo {author} {\bibfnamefont {S.}~\bibnamefont {Aunsmo}}\ and\ \bibinfo {author} {\bibfnamefont {J.}~\bibnamefont {Linder}},\ }\bibfield  {title} {\bibinfo {title} {Converting a triplet cooper pair supercurrent into a spin signal},\ }\href {https://doi.org/10.1103/PhysRevB.109.024503} {\bibfield  {journal} {\bibinfo  {journal} {Phys. Rev. B}\ }\textbf {\bibinfo {volume} {109}},\ \bibinfo {pages} {024503} (\bibinfo {year} {2024})}\BibitemShut {NoStop}%
\bibitem [{\citenamefont {Edelstein}(1995)}]{Edelstein1995}%
  \BibitemOpen
  \bibfield  {author} {\bibinfo {author} {\bibfnamefont {V.~M.}\ \bibnamefont {Edelstein}},\ }\bibfield  {title} {\bibinfo {title} {Magnetoelectric effect in polar superconductors},\ }\href {https://doi.org/10.1103/PhysRevLett.75.2004} {\bibfield  {journal} {\bibinfo  {journal} {Phys. Rev. Lett.}\ }\textbf {\bibinfo {volume} {75}},\ \bibinfo {pages} {2004} (\bibinfo {year} {1995})}\BibitemShut {NoStop}%
\bibitem [{\citenamefont {Edelstein}(2005)}]{Edelstein2005}%
  \BibitemOpen
  \bibfield  {author} {\bibinfo {author} {\bibfnamefont {V.~M.}\ \bibnamefont {Edelstein}},\ }\bibfield  {title} {\bibinfo {title} {Magnetoelectric effect in dirty superconductors with broken mirror symmetry},\ }\href {https://doi.org/10.1103/PhysRevB.72.172501} {\bibfield  {journal} {\bibinfo  {journal} {Phys. Rev. B}\ }\textbf {\bibinfo {volume} {72}},\ \bibinfo {pages} {172501} (\bibinfo {year} {2005})}\BibitemShut {NoStop}%
\bibitem [{\citenamefont {Ili\ifmmode~\acute{c}\else \'{c}\fi{}}\ \emph {et~al.}(2020)\citenamefont {Ili\ifmmode~\acute{c}\else \'{c}\fi{}}, \citenamefont {Tokatly},\ and\ \citenamefont {Bergeret}}]{Ilic2020}%
  \BibitemOpen
  \bibfield  {author} {\bibinfo {author} {\bibfnamefont {S.}~\bibnamefont {Ili\ifmmode~\acute{c}\else \'{c}\fi{}}}, \bibinfo {author} {\bibfnamefont {I.~V.}\ \bibnamefont {Tokatly}},\ and\ \bibinfo {author} {\bibfnamefont {F.~S.}\ \bibnamefont {Bergeret}},\ }\bibfield  {title} {\bibinfo {title} {Unified description of spin transport, weak antilocalization, and triplet superconductivity in systems with spin-orbit coupling},\ }\href {https://doi.org/10.1103/PhysRevB.102.235430} {\bibfield  {journal} {\bibinfo  {journal} {Phys. Rev. B}\ }\textbf {\bibinfo {volume} {102}},\ \bibinfo {pages} {235430} (\bibinfo {year} {2020})}\BibitemShut {NoStop}%
\bibitem [{\citenamefont {Zhu}\ \emph {et~al.}(2004)\citenamefont {Zhu}, \citenamefont {Nussinov}, \citenamefont {Shnirman},\ and\ \citenamefont {Balatsky}}]{Zhu2004}%
  \BibitemOpen
  \bibfield  {author} {\bibinfo {author} {\bibfnamefont {J.-X.}\ \bibnamefont {Zhu}}, \bibinfo {author} {\bibfnamefont {Z.}~\bibnamefont {Nussinov}}, \bibinfo {author} {\bibfnamefont {A.}~\bibnamefont {Shnirman}},\ and\ \bibinfo {author} {\bibfnamefont {A.~V.}\ \bibnamefont {Balatsky}},\ }\bibfield  {title} {\bibinfo {title} {Novel spin dynamics in a josephson junction},\ }\href {https://doi.org/10.1103/PhysRevLett.92.107001} {\bibfield  {journal} {\bibinfo  {journal} {Phys. Rev. Lett.}\ }\textbf {\bibinfo {volume} {92}},\ \bibinfo {pages} {107001} (\bibinfo {year} {2004})}\BibitemShut {NoStop}%
\bibitem [{\citenamefont {Nussinov}\ \emph {et~al.}(2005)\citenamefont {Nussinov}, \citenamefont {Shnirman}, \citenamefont {Arovas}, \citenamefont {Balatsky},\ and\ \citenamefont {Zhu}}]{Zhu2005}%
  \BibitemOpen
  \bibfield  {author} {\bibinfo {author} {\bibfnamefont {Z.}~\bibnamefont {Nussinov}}, \bibinfo {author} {\bibfnamefont {A.}~\bibnamefont {Shnirman}}, \bibinfo {author} {\bibfnamefont {D.~P.}\ \bibnamefont {Arovas}}, \bibinfo {author} {\bibfnamefont {A.~V.}\ \bibnamefont {Balatsky}},\ and\ \bibinfo {author} {\bibfnamefont {J.~X.}\ \bibnamefont {Zhu}},\ }\bibfield  {title} {\bibinfo {title} {Spin and spin-wave dynamics in josephson junctions},\ }\href {https://doi.org/10.1103/PhysRevB.71.214520} {\bibfield  {journal} {\bibinfo  {journal} {Phys. Rev. B}\ }\textbf {\bibinfo {volume} {71}},\ \bibinfo {pages} {214520} (\bibinfo {year} {2005})}\BibitemShut {NoStop}%
\bibitem [{\citenamefont {Holmqvist}\ \emph {et~al.}(2011)\citenamefont {Holmqvist}, \citenamefont {Teber},\ and\ \citenamefont {Fogelstr\"om}}]{Holmqvist2011}%
  \BibitemOpen
  \bibfield  {author} {\bibinfo {author} {\bibfnamefont {C.}~\bibnamefont {Holmqvist}}, \bibinfo {author} {\bibfnamefont {S.}~\bibnamefont {Teber}},\ and\ \bibinfo {author} {\bibfnamefont {M.}~\bibnamefont {Fogelstr\"om}},\ }\bibfield  {title} {\bibinfo {title} {Nonequilibrium effects in a josephson junction coupled to a precessing spin},\ }\href {https://doi.org/10.1103/PhysRevB.83.104521} {\bibfield  {journal} {\bibinfo  {journal} {Phys. Rev. B}\ }\textbf {\bibinfo {volume} {83}},\ \bibinfo {pages} {104521} (\bibinfo {year} {2011})}\BibitemShut {NoStop}%
\bibitem [{\citenamefont {Linder}\ and\ \citenamefont {Yokoyama}(2011)}]{Linder2011}%
  \BibitemOpen
  \bibfield  {author} {\bibinfo {author} {\bibfnamefont {J.}~\bibnamefont {Linder}}\ and\ \bibinfo {author} {\bibfnamefont {T.}~\bibnamefont {Yokoyama}},\ }\bibfield  {title} {\bibinfo {title} {Supercurrent-induced magnetization dynamics in a josephson junction with two misaligned ferromagnetic layers},\ }\href {https://doi.org/10.1103/PhysRevB.83.012501} {\bibfield  {journal} {\bibinfo  {journal} {Phys. Rev. B}\ }\textbf {\bibinfo {volume} {83}},\ \bibinfo {pages} {012501} (\bibinfo {year} {2011})}\BibitemShut {NoStop}%
\bibitem [{\citenamefont {Halterman}\ and\ \citenamefont {Alidoust}(2016)}]{Halterman2016}%
  \BibitemOpen
  \bibfield  {author} {\bibinfo {author} {\bibfnamefont {K.}~\bibnamefont {Halterman}}\ and\ \bibinfo {author} {\bibfnamefont {M.}~\bibnamefont {Alidoust}},\ }\bibfield  {title} {\bibinfo {title} {Josephson currents and spin-transfer torques in ballistic sfsfs nanojunctions},\ }\href {https://doi.org/10.1088/0953-2048/29/5/055007} {\bibfield  {journal} {\bibinfo  {journal} {Superconductor Science and Technology}\ }\textbf {\bibinfo {volume} {29}},\ \bibinfo {pages} {055007} (\bibinfo {year} {2016})}\BibitemShut {NoStop}%
\bibitem [{\citenamefont {Kulagina}\ and\ \citenamefont {Linder}(2014)}]{Kulagina2014}%
  \BibitemOpen
  \bibfield  {author} {\bibinfo {author} {\bibfnamefont {I.}~\bibnamefont {Kulagina}}\ and\ \bibinfo {author} {\bibfnamefont {J.}~\bibnamefont {Linder}},\ }\bibfield  {title} {\bibinfo {title} {Spin supercurrent, magnetization dynamics, and $\ensuremath{\varphi}$-state in spin-textured josephson junctions},\ }\href {https://doi.org/10.1103/PhysRevB.90.054504} {\bibfield  {journal} {\bibinfo  {journal} {Phys. Rev. B}\ }\textbf {\bibinfo {volume} {90}},\ \bibinfo {pages} {054504} (\bibinfo {year} {2014})}\BibitemShut {NoStop}%
\bibitem [{\citenamefont {Linder}\ \emph {et~al.}(2012)\citenamefont {Linder}, \citenamefont {Brataas}, \citenamefont {Shomali},\ and\ \citenamefont {Zareyan}}]{Linder2012}%
  \BibitemOpen
  \bibfield  {author} {\bibinfo {author} {\bibfnamefont {J.}~\bibnamefont {Linder}}, \bibinfo {author} {\bibfnamefont {A.}~\bibnamefont {Brataas}}, \bibinfo {author} {\bibfnamefont {Z.}~\bibnamefont {Shomali}},\ and\ \bibinfo {author} {\bibfnamefont {M.}~\bibnamefont {Zareyan}},\ }\bibfield  {title} {\bibinfo {title} {Spin-transfer and exchange torques in ferromagnetic superconductors},\ }\href {https://doi.org/10.1103/PhysRevLett.109.237206} {\bibfield  {journal} {\bibinfo  {journal} {Phys. Rev. Lett.}\ }\textbf {\bibinfo {volume} {109}},\ \bibinfo {pages} {237206} (\bibinfo {year} {2012})}\BibitemShut {NoStop}%
\bibitem [{\citenamefont {Takashima}\ \emph {et~al.}(2017)\citenamefont {Takashima}, \citenamefont {Fujimoto},\ and\ \citenamefont {Yokoyama}}]{Takashima2017}%
  \BibitemOpen
  \bibfield  {author} {\bibinfo {author} {\bibfnamefont {R.}~\bibnamefont {Takashima}}, \bibinfo {author} {\bibfnamefont {S.}~\bibnamefont {Fujimoto}},\ and\ \bibinfo {author} {\bibfnamefont {T.}~\bibnamefont {Yokoyama}},\ }\bibfield  {title} {\bibinfo {title} {Adiabatic and nonadiabatic spin torques induced by a spin-triplet supercurrent},\ }\href {https://doi.org/10.1103/PhysRevB.96.121203} {\bibfield  {journal} {\bibinfo  {journal} {Phys. Rev. B}\ }\textbf {\bibinfo {volume} {96}},\ \bibinfo {pages} {121203} (\bibinfo {year} {2017})}\BibitemShut {NoStop}%
\bibitem [{\citenamefont {Shrivastava}\ and\ \citenamefont {Ramgopal~Rao}(2021)}]{Shrivastava2021}%
  \BibitemOpen
  \bibfield  {author} {\bibinfo {author} {\bibfnamefont {M.}~\bibnamefont {Shrivastava}}\ and\ \bibinfo {author} {\bibfnamefont {V.}~\bibnamefont {Ramgopal~Rao}},\ }\bibfield  {title} {\bibinfo {title} {A roadmap for disruptive applications and heterogeneous integration using two-dimensional materials: State-of-the-art and technological challenges},\ }\href {https://doi.org/10.1021/acs.nanolett.1c00729} {\bibfield  {journal} {\bibinfo  {journal} {Nano Letters}\ }\textbf {\bibinfo {volume} {21}},\ \bibinfo {pages} {6359} (\bibinfo {year} {2021})}\BibitemShut {NoStop}%
\bibitem [{\citenamefont {Geim}\ and\ \citenamefont {Grigorieva}(2013)}]{Geim2013}%
  \BibitemOpen
  \bibfield  {author} {\bibinfo {author} {\bibfnamefont {A.~K.}\ \bibnamefont {Geim}}\ and\ \bibinfo {author} {\bibfnamefont {I.~V.}\ \bibnamefont {Grigorieva}},\ }\bibfield  {title} {\bibinfo {title} {Van der waals heterostructures},\ }\href {https://doi.org/10.1038/nature12385} {\bibfield  {journal} {\bibinfo  {journal} {Nature}\ }\textbf {\bibinfo {volume} {499}},\ \bibinfo {pages} {419} (\bibinfo {year} {2013})}\BibitemShut {NoStop}%
\bibitem [{\citenamefont {Khokhriakov}\ \emph {et~al.}(2020)\citenamefont {Khokhriakov}, \citenamefont {Hoque}, \citenamefont {Karpiak},\ and\ \citenamefont {Dash}}]{Khokhriakov2020}%
  \BibitemOpen
  \bibfield  {author} {\bibinfo {author} {\bibfnamefont {D.}~\bibnamefont {Khokhriakov}}, \bibinfo {author} {\bibfnamefont {A.~M.}\ \bibnamefont {Hoque}}, \bibinfo {author} {\bibfnamefont {B.}~\bibnamefont {Karpiak}},\ and\ \bibinfo {author} {\bibfnamefont {S.~P.}\ \bibnamefont {Dash}},\ }\bibfield  {title} {\bibinfo {title} {Gate-tunable spin-galvanic effect in graphene-topological insulator van der waals heterostructures at room temperature},\ }\href {https://doi.org/10.1038/s41467-020-17481-1} {\bibfield  {journal} {\bibinfo  {journal} {Nature Communications}\ }\textbf {\bibinfo {volume} {11}},\ \bibinfo {pages} {3657} (\bibinfo {year} {2020})}\BibitemShut {NoStop}%
\bibitem [{\citenamefont {Lin}\ \emph {et~al.}(2019)\citenamefont {Lin}, \citenamefont {Yang}, \citenamefont {Wang},\ and\ \citenamefont {Zhao}}]{Lin2019}%
  \BibitemOpen
  \bibfield  {author} {\bibinfo {author} {\bibfnamefont {X.}~\bibnamefont {Lin}}, \bibinfo {author} {\bibfnamefont {W.}~\bibnamefont {Yang}}, \bibinfo {author} {\bibfnamefont {K.~L.}\ \bibnamefont {Wang}},\ and\ \bibinfo {author} {\bibfnamefont {W.}~\bibnamefont {Zhao}},\ }\bibfield  {title} {\bibinfo {title} {Two-dimensional spintronics for low-power electronics},\ }\href {https://doi.org/10.1038/s41928-019-0273-7} {\bibfield  {journal} {\bibinfo  {journal} {Nature Electronics}\ }\textbf {\bibinfo {volume} {2}},\ \bibinfo {pages} {274} (\bibinfo {year} {2019})}\BibitemShut {NoStop}%
\bibitem [{\citenamefont {Guimar{\~a}es}\ \emph {et~al.}(2018)\citenamefont {Guimar{\~a}es}, \citenamefont {Stiehl}, \citenamefont {MacNeill}, \citenamefont {Reynolds},\ and\ \citenamefont {Ralph}}]{Guimaraes2018}%
  \BibitemOpen
  \bibfield  {author} {\bibinfo {author} {\bibfnamefont {M.~H.~D.}\ \bibnamefont {Guimar{\~a}es}}, \bibinfo {author} {\bibfnamefont {G.~M.}\ \bibnamefont {Stiehl}}, \bibinfo {author} {\bibfnamefont {D.}~\bibnamefont {MacNeill}}, \bibinfo {author} {\bibfnamefont {N.~D.}\ \bibnamefont {Reynolds}},\ and\ \bibinfo {author} {\bibfnamefont {D.~C.}\ \bibnamefont {Ralph}},\ }\bibfield  {title} {\bibinfo {title} {Spin--orbit torques in nbse2/permalloy bilayers},\ }\href {https://doi.org/10.1021/acs.nanolett.7b04993} {\bibfield  {journal} {\bibinfo  {journal} {Nano Letters}\ }\textbf {\bibinfo {volume} {18}},\ \bibinfo {pages} {1311} (\bibinfo {year} {2018})}\BibitemShut {NoStop}%
\bibitem [{\citenamefont {Shi}\ \emph {et~al.}(2019)\citenamefont {Shi}, \citenamefont {Liang}, \citenamefont {Zhu}, \citenamefont {Cai}, \citenamefont {Pollard}, \citenamefont {Wang}, \citenamefont {Wang}, \citenamefont {Wang}, \citenamefont {He}, \citenamefont {Yu}, \citenamefont {Eda}, \citenamefont {Liang},\ and\ \citenamefont {Yang}}]{Shi2019}%
  \BibitemOpen
  \bibfield  {author} {\bibinfo {author} {\bibfnamefont {S.}~\bibnamefont {Shi}}, \bibinfo {author} {\bibfnamefont {S.}~\bibnamefont {Liang}}, \bibinfo {author} {\bibfnamefont {Z.}~\bibnamefont {Zhu}}, \bibinfo {author} {\bibfnamefont {K.}~\bibnamefont {Cai}}, \bibinfo {author} {\bibfnamefont {S.~D.}\ \bibnamefont {Pollard}}, \bibinfo {author} {\bibfnamefont {Y.}~\bibnamefont {Wang}}, \bibinfo {author} {\bibfnamefont {J.}~\bibnamefont {Wang}}, \bibinfo {author} {\bibfnamefont {Q.}~\bibnamefont {Wang}}, \bibinfo {author} {\bibfnamefont {P.}~\bibnamefont {He}}, \bibinfo {author} {\bibfnamefont {J.}~\bibnamefont {Yu}}, \bibinfo {author} {\bibfnamefont {G.}~\bibnamefont {Eda}}, \bibinfo {author} {\bibfnamefont {G.}~\bibnamefont {Liang}},\ and\ \bibinfo {author} {\bibfnamefont {H.}~\bibnamefont {Yang}},\ }\bibfield  {title} {\bibinfo {title} {All-electric magnetization switching and dzyaloshinskii--moriya interaction in wte2/ferromagnet heterostructures},\ }\href {https://doi.org/10.1038/s41565-019-0525-8}
  {\bibfield  {journal} {\bibinfo  {journal} {Nature Nanotechnology}\ }\textbf {\bibinfo {volume} {14}},\ \bibinfo {pages} {945} (\bibinfo {year} {2019})}\BibitemShut {NoStop}%
\bibitem [{\citenamefont {MacNeill}\ \emph {et~al.}(2017)\citenamefont {MacNeill}, \citenamefont {Stiehl}, \citenamefont {Guimaraes}, \citenamefont {Buhrman}, \citenamefont {Park},\ and\ \citenamefont {Ralph}}]{MacNeill2017}%
  \BibitemOpen
  \bibfield  {author} {\bibinfo {author} {\bibfnamefont {D.}~\bibnamefont {MacNeill}}, \bibinfo {author} {\bibfnamefont {G.~M.}\ \bibnamefont {Stiehl}}, \bibinfo {author} {\bibfnamefont {M.~H.~D.}\ \bibnamefont {Guimaraes}}, \bibinfo {author} {\bibfnamefont {R.~A.}\ \bibnamefont {Buhrman}}, \bibinfo {author} {\bibfnamefont {J.}~\bibnamefont {Park}},\ and\ \bibinfo {author} {\bibfnamefont {D.~C.}\ \bibnamefont {Ralph}},\ }\bibfield  {title} {\bibinfo {title} {Control of spin--orbit torques through crystal symmetry in wte2/ferromagnet bilayers},\ }\href {https://doi.org/10.1038/nphys3933} {\bibfield  {journal} {\bibinfo  {journal} {Nature Physics}\ }\textbf {\bibinfo {volume} {13}},\ \bibinfo {pages} {300} (\bibinfo {year} {2017})}\BibitemShut {NoStop}%
\bibitem [{\citenamefont {Zhao}\ \emph {et~al.}(2020{\natexlab{a}})\citenamefont {Zhao}, \citenamefont {Karpiak}, \citenamefont {Khokhriakov}, \citenamefont {Johansson}, \citenamefont {Hoque}, \citenamefont {Xu}, \citenamefont {Jiang}, \citenamefont {Mertig},\ and\ \citenamefont {Dash}}]{Zhao2020}%
  \BibitemOpen
  \bibfield  {author} {\bibinfo {author} {\bibfnamefont {B.}~\bibnamefont {Zhao}}, \bibinfo {author} {\bibfnamefont {B.}~\bibnamefont {Karpiak}}, \bibinfo {author} {\bibfnamefont {D.}~\bibnamefont {Khokhriakov}}, \bibinfo {author} {\bibfnamefont {A.}~\bibnamefont {Johansson}}, \bibinfo {author} {\bibfnamefont {A.~M.}\ \bibnamefont {Hoque}}, \bibinfo {author} {\bibfnamefont {X.}~\bibnamefont {Xu}}, \bibinfo {author} {\bibfnamefont {Y.}~\bibnamefont {Jiang}}, \bibinfo {author} {\bibfnamefont {I.}~\bibnamefont {Mertig}},\ and\ \bibinfo {author} {\bibfnamefont {S.~P.}\ \bibnamefont {Dash}},\ }\bibfield  {title} {\bibinfo {title} {Unconventional charge--spin conversion in weyl-semimetal wte2},\ }\href {https://doi.org/10.1002/adma.202000818} {\bibfield  {journal} {\bibinfo  {journal} {Advanced Materials}\ }\textbf {\bibinfo {volume} {32}},\ \bibinfo {pages} {2000818} (\bibinfo {year} {2020}{\natexlab{a}})}\BibitemShut {NoStop}%
\bibitem [{\citenamefont {Stiehl}\ \emph {et~al.}(2019)\citenamefont {Stiehl}, \citenamefont {Li}, \citenamefont {Gupta}, \citenamefont {Baggari}, \citenamefont {Jiang}, \citenamefont {Xie}, \citenamefont {Kourkoutis}, \citenamefont {Mak}, \citenamefont {Shan}, \citenamefont {Buhrman},\ and\ \citenamefont {Ralph}}]{Stiehl2019}%
  \BibitemOpen
  \bibfield  {author} {\bibinfo {author} {\bibfnamefont {G.~M.}\ \bibnamefont {Stiehl}}, \bibinfo {author} {\bibfnamefont {R.}~\bibnamefont {Li}}, \bibinfo {author} {\bibfnamefont {V.}~\bibnamefont {Gupta}}, \bibinfo {author} {\bibfnamefont {I.~E.}\ \bibnamefont {Baggari}}, \bibinfo {author} {\bibfnamefont {S.}~\bibnamefont {Jiang}}, \bibinfo {author} {\bibfnamefont {H.}~\bibnamefont {Xie}}, \bibinfo {author} {\bibfnamefont {L.~F.}\ \bibnamefont {Kourkoutis}}, \bibinfo {author} {\bibfnamefont {K.~F.}\ \bibnamefont {Mak}}, \bibinfo {author} {\bibfnamefont {J.}~\bibnamefont {Shan}}, \bibinfo {author} {\bibfnamefont {R.~A.}\ \bibnamefont {Buhrman}},\ and\ \bibinfo {author} {\bibfnamefont {D.~C.}\ \bibnamefont {Ralph}},\ }\bibfield  {title} {\bibinfo {title} {Layer-dependent spin-orbit torques generated by the centrosymmetric transition metal dichalcogenide $\ensuremath{\beta}\ensuremath{-}{\mathrm{mote}}_{2}$},\ }\href {https://doi.org/10.1103/PhysRevB.100.184402} {\bibfield  {journal} {\bibinfo  {journal}
  {Phys. Rev. B}\ }\textbf {\bibinfo {volume} {100}},\ \bibinfo {pages} {184402} (\bibinfo {year} {2019})}\BibitemShut {NoStop}%
\bibitem [{\citenamefont {Safeer}\ \emph {et~al.}(2019)\citenamefont {Safeer}, \citenamefont {Ontoso}, \citenamefont {Ingla-Ayn{\'e}s}, \citenamefont {Herling}, \citenamefont {Pham}, \citenamefont {Kurzmann}, \citenamefont {Ensslin}, \citenamefont {Chuvilin}, \citenamefont {Robredo}, \citenamefont {Vergniory}, \citenamefont {de~Juan}, \citenamefont {Hueso}, \citenamefont {Calvo},\ and\ \citenamefont {Casanova}}]{Safeer2019}%
  \BibitemOpen
  \bibfield  {author} {\bibinfo {author} {\bibfnamefont {C.~K.}\ \bibnamefont {Safeer}}, \bibinfo {author} {\bibfnamefont {N.}~\bibnamefont {Ontoso}}, \bibinfo {author} {\bibfnamefont {J.}~\bibnamefont {Ingla-Ayn{\'e}s}}, \bibinfo {author} {\bibfnamefont {F.}~\bibnamefont {Herling}}, \bibinfo {author} {\bibfnamefont {V.~T.}\ \bibnamefont {Pham}}, \bibinfo {author} {\bibfnamefont {A.}~\bibnamefont {Kurzmann}}, \bibinfo {author} {\bibfnamefont {K.}~\bibnamefont {Ensslin}}, \bibinfo {author} {\bibfnamefont {A.}~\bibnamefont {Chuvilin}}, \bibinfo {author} {\bibfnamefont {I.}~\bibnamefont {Robredo}}, \bibinfo {author} {\bibfnamefont {M.~G.}\ \bibnamefont {Vergniory}}, \bibinfo {author} {\bibfnamefont {F.}~\bibnamefont {de~Juan}}, \bibinfo {author} {\bibfnamefont {L.~E.}\ \bibnamefont {Hueso}}, \bibinfo {author} {\bibfnamefont {M.~R.}\ \bibnamefont {Calvo}},\ and\ \bibinfo {author} {\bibfnamefont {F.}~\bibnamefont {Casanova}},\ }\bibfield  {title} {\bibinfo {title} {Large multidirectional spin-to-charge conversion
  in low-symmetry semimetal mote2 at room temperature},\ }\href {https://doi.org/10.1021/acs.nanolett.9b03485} {\bibfield  {journal} {\bibinfo  {journal} {Nano Letters}\ }\textbf {\bibinfo {volume} {19}},\ \bibinfo {pages} {8758} (\bibinfo {year} {2019})}\BibitemShut {NoStop}%
\bibitem [{\citenamefont {Zhao}\ \emph {et~al.}(2020{\natexlab{b}})\citenamefont {Zhao}, \citenamefont {Khokhriakov}, \citenamefont {Zhang}, \citenamefont {Fu}, \citenamefont {Karpiak}, \citenamefont {Hoque}, \citenamefont {Xu}, \citenamefont {Jiang}, \citenamefont {Yan},\ and\ \citenamefont {Dash}}]{Zhao2020_WTe2}%
  \BibitemOpen
  \bibfield  {author} {\bibinfo {author} {\bibfnamefont {B.}~\bibnamefont {Zhao}}, \bibinfo {author} {\bibfnamefont {D.}~\bibnamefont {Khokhriakov}}, \bibinfo {author} {\bibfnamefont {Y.}~\bibnamefont {Zhang}}, \bibinfo {author} {\bibfnamefont {H.}~\bibnamefont {Fu}}, \bibinfo {author} {\bibfnamefont {B.}~\bibnamefont {Karpiak}}, \bibinfo {author} {\bibfnamefont {A.~M.}\ \bibnamefont {Hoque}}, \bibinfo {author} {\bibfnamefont {X.}~\bibnamefont {Xu}}, \bibinfo {author} {\bibfnamefont {Y.}~\bibnamefont {Jiang}}, \bibinfo {author} {\bibfnamefont {B.}~\bibnamefont {Yan}},\ and\ \bibinfo {author} {\bibfnamefont {S.~P.}\ \bibnamefont {Dash}},\ }\bibfield  {title} {\bibinfo {title} {Observation of charge to spin conversion in weyl semimetal ${\mathrm{wte}}_{2}$ at room temperature},\ }\href {https://doi.org/10.1103/PhysRevResearch.2.013286} {\bibfield  {journal} {\bibinfo  {journal} {Phys. Rev. Res.}\ }\textbf {\bibinfo {volume} {2}},\ \bibinfo {pages} {013286} (\bibinfo {year} {2020}{\natexlab{b}})}\BibitemShut
  {NoStop}%
\bibitem [{\citenamefont {Ghiasi}\ \emph {et~al.}(2019)\citenamefont {Ghiasi}, \citenamefont {Kaverzin}, \citenamefont {Blah},\ and\ \citenamefont {van Wees}}]{Ghiasi2019}%
  \BibitemOpen
  \bibfield  {author} {\bibinfo {author} {\bibfnamefont {T.~S.}\ \bibnamefont {Ghiasi}}, \bibinfo {author} {\bibfnamefont {A.~A.}\ \bibnamefont {Kaverzin}}, \bibinfo {author} {\bibfnamefont {P.~J.}\ \bibnamefont {Blah}},\ and\ \bibinfo {author} {\bibfnamefont {B.~J.}\ \bibnamefont {van Wees}},\ }\bibfield  {title} {\bibinfo {title} {Charge-to-spin conversion by the rashba--edelstein effect in two-dimensional van der waals heterostructures up to room temperature},\ }\href {https://doi.org/10.1021/acs.nanolett.9b01611} {\bibfield  {journal} {\bibinfo  {journal} {Nano Letters}\ }\textbf {\bibinfo {volume} {19}},\ \bibinfo {pages} {5959} (\bibinfo {year} {2019})}\BibitemShut {NoStop}%
\bibitem [{\citenamefont {Hoque}\ \emph {et~al.}(2022)\citenamefont {Hoque}, \citenamefont {Zhao}, \citenamefont {Khokhriakov}, \citenamefont {Muduli},\ and\ \citenamefont {Dash}}]{Hoque2022}%
  \BibitemOpen
  \bibfield  {author} {\bibinfo {author} {\bibfnamefont {A.~M.}\ \bibnamefont {Hoque}}, \bibinfo {author} {\bibfnamefont {B.}~\bibnamefont {Zhao}}, \bibinfo {author} {\bibfnamefont {D.}~\bibnamefont {Khokhriakov}}, \bibinfo {author} {\bibfnamefont {P.}~\bibnamefont {Muduli}},\ and\ \bibinfo {author} {\bibfnamefont {S.~P.}\ \bibnamefont {Dash}},\ }\bibfield  {title} {\bibinfo {title} {Charge to spin conversion in van der waals metal nbse2},\ }\href {https://doi.org/10.1063/5.0121577} {\bibfield  {journal} {\bibinfo  {journal} {Applied Physics Letters}\ }\textbf {\bibinfo {volume} {121}},\ \bibinfo {pages} {242404} (\bibinfo {year} {2022})}\BibitemShut {NoStop}%
\bibitem [{\citenamefont {Bobkov}\ \emph {et~al.}(2024)\citenamefont {Bobkov}, \citenamefont {Bokai}, \citenamefont {Otrokov}, \citenamefont {Bobkov},\ and\ \citenamefont {Bobkova}}]{Bobkov2024_vdW}%
  \BibitemOpen
  \bibfield  {author} {\bibinfo {author} {\bibfnamefont {G.~A.}\ \bibnamefont {Bobkov}}, \bibinfo {author} {\bibfnamefont {K.~A.}\ \bibnamefont {Bokai}}, \bibinfo {author} {\bibfnamefont {M.~M.}\ \bibnamefont {Otrokov}}, \bibinfo {author} {\bibfnamefont {A.~M.}\ \bibnamefont {Bobkov}},\ and\ \bibinfo {author} {\bibfnamefont {I.~V.}\ \bibnamefont {Bobkova}},\ }\href@noop {} {\bibinfo {title} {Gate-controlled superconducting proximity effect of superconductor/ferromagnet van der waals heterostructures}} (\bibinfo {year} {2024}),\ \Eprint {https://arxiv.org/abs/2405.07575} {arXiv:2405.07575 [cond-mat.supr-con]} \BibitemShut {NoStop}%
\bibitem [{\citenamefont {Aikebaier}\ \emph {et~al.}(2022)\citenamefont {Aikebaier}, \citenamefont {Heikkil\"a},\ and\ \citenamefont {Lado}}]{Aikebaier2022}%
  \BibitemOpen
  \bibfield  {author} {\bibinfo {author} {\bibfnamefont {F.}~\bibnamefont {Aikebaier}}, \bibinfo {author} {\bibfnamefont {T.~T.}\ \bibnamefont {Heikkil\"a}},\ and\ \bibinfo {author} {\bibfnamefont {J.~L.}\ \bibnamefont {Lado}},\ }\bibfield  {title} {\bibinfo {title} {Controlling magnetism through ising superconductivity in magnetic van der waals heterostructures},\ }\href {https://doi.org/10.1103/PhysRevB.105.054506} {\bibfield  {journal} {\bibinfo  {journal} {Phys. Rev. B}\ }\textbf {\bibinfo {volume} {105}},\ \bibinfo {pages} {054506} (\bibinfo {year} {2022})}\BibitemShut {NoStop}%
\bibitem [{\citenamefont {Bobkov}\ \emph {et~al.}(2022)\citenamefont {Bobkov}, \citenamefont {Bobkova}, \citenamefont {Bobkov},\ and\ \citenamefont {Kamra}}]{Bobkov2022}%
  \BibitemOpen
  \bibfield  {author} {\bibinfo {author} {\bibfnamefont {G.~A.}\ \bibnamefont {Bobkov}}, \bibinfo {author} {\bibfnamefont {I.~V.}\ \bibnamefont {Bobkova}}, \bibinfo {author} {\bibfnamefont {A.~M.}\ \bibnamefont {Bobkov}},\ and\ \bibinfo {author} {\bibfnamefont {A.}~\bibnamefont {Kamra}},\ }\bibfield  {title} {\bibinfo {title} {N\'eel proximity effect at antiferromagnet/superconductor interfaces},\ }\href {https://doi.org/10.1103/PhysRevB.106.144512} {\bibfield  {journal} {\bibinfo  {journal} {Phys. Rev. B}\ }\textbf {\bibinfo {volume} {106}},\ \bibinfo {pages} {144512} (\bibinfo {year} {2022})}\BibitemShut {NoStop}%
\bibitem [{\citenamefont {Wickramaratne}\ \emph {et~al.}(2020)\citenamefont {Wickramaratne}, \citenamefont {Khmelevskyi}, \citenamefont {Agterberg},\ and\ \citenamefont {Mazin}}]{Wickramaratne2020}%
  \BibitemOpen
  \bibfield  {author} {\bibinfo {author} {\bibfnamefont {D.}~\bibnamefont {Wickramaratne}}, \bibinfo {author} {\bibfnamefont {S.}~\bibnamefont {Khmelevskyi}}, \bibinfo {author} {\bibfnamefont {D.~F.}\ \bibnamefont {Agterberg}},\ and\ \bibinfo {author} {\bibfnamefont {I.~I.}\ \bibnamefont {Mazin}},\ }\bibfield  {title} {\bibinfo {title} {Ising superconductivity and magnetism in ${\mathrm{nbse}}_{2}$},\ }\href {https://doi.org/10.1103/PhysRevX.10.041003} {\bibfield  {journal} {\bibinfo  {journal} {Phys. Rev. X}\ }\textbf {\bibinfo {volume} {10}},\ \bibinfo {pages} {041003} (\bibinfo {year} {2020})}\BibitemShut {NoStop}%
\bibitem [{\citenamefont {Serene}\ and\ \citenamefont {Rainer}(1983)}]{Serene1983}%
  \BibitemOpen
  \bibfield  {author} {\bibinfo {author} {\bibfnamefont {J.~W.}\ \bibnamefont {Serene}}\ and\ \bibinfo {author} {\bibfnamefont {D.}~\bibnamefont {Rainer}},\ }\bibfield  {title} {\bibinfo {title} {The quasiclassical approach to superfluid 3he},\ }\href {https://www.sciencedirect.com/science/article/pii/0370157383900510} {\bibfield  {journal} {\bibinfo  {journal} {Physics Reports}\ }\textbf {\bibinfo {volume} {101}},\ \bibinfo {pages} {221} (\bibinfo {year} {1983})}\BibitemShut {NoStop}%
\bibitem [{\citenamefont {Sigrist}\ and\ \citenamefont {Ueda}(1991)}]{Sigrist1991}%
  \BibitemOpen
  \bibfield  {author} {\bibinfo {author} {\bibfnamefont {M.}~\bibnamefont {Sigrist}}\ and\ \bibinfo {author} {\bibfnamefont {K.}~\bibnamefont {Ueda}},\ }\bibfield  {title} {\bibinfo {title} {Phenomenological theory of unconventional superconductivity},\ }\href {https://doi.org/10.1103/RevModPhys.63.239} {\bibfield  {journal} {\bibinfo  {journal} {Rev. Mod. Phys.}\ }\textbf {\bibinfo {volume} {63}},\ \bibinfo {pages} {239} (\bibinfo {year} {1991})}\BibitemShut {NoStop}%
\bibitem [{\citenamefont {M\"ockli}\ and\ \citenamefont {Khodas}(2020)}]{Mockli2020}%
  \BibitemOpen
  \bibfield  {author} {\bibinfo {author} {\bibfnamefont {D.}~\bibnamefont {M\"ockli}}\ and\ \bibinfo {author} {\bibfnamefont {M.}~\bibnamefont {Khodas}},\ }\bibfield  {title} {\bibinfo {title} {Ising superconductors: Interplay of magnetic field, triplet channels, and disorder},\ }\href {https://doi.org/10.1103/PhysRevB.101.014510} {\bibfield  {journal} {\bibinfo  {journal} {Phys. Rev. B}\ }\textbf {\bibinfo {volume} {101}},\ \bibinfo {pages} {014510} (\bibinfo {year} {2020})}\BibitemShut {NoStop}%
\bibitem [{\citenamefont {Xi}\ \emph {et~al.}(2016)\citenamefont {Xi}, \citenamefont {Wang}, \citenamefont {Zhao}, \citenamefont {Park}, \citenamefont {Law}, \citenamefont {Berger}, \citenamefont {Forr{\'o}}, \citenamefont {Shan},\ and\ \citenamefont {Mak}}]{Xi2016}%
  \BibitemOpen
  \bibfield  {author} {\bibinfo {author} {\bibfnamefont {X.}~\bibnamefont {Xi}}, \bibinfo {author} {\bibfnamefont {Z.}~\bibnamefont {Wang}}, \bibinfo {author} {\bibfnamefont {W.}~\bibnamefont {Zhao}}, \bibinfo {author} {\bibfnamefont {J.-H.}\ \bibnamefont {Park}}, \bibinfo {author} {\bibfnamefont {K.~T.}\ \bibnamefont {Law}}, \bibinfo {author} {\bibfnamefont {H.}~\bibnamefont {Berger}}, \bibinfo {author} {\bibfnamefont {L.}~\bibnamefont {Forr{\'o}}}, \bibinfo {author} {\bibfnamefont {J.}~\bibnamefont {Shan}},\ and\ \bibinfo {author} {\bibfnamefont {K.~F.}\ \bibnamefont {Mak}},\ }\bibfield  {title} {\bibinfo {title} {Ising pairing in superconducting nbse2 atomic layers},\ }\href {https://doi.org/10.1038/nphys3538} {\bibfield  {journal} {\bibinfo  {journal} {Nature Physics}\ }\textbf {\bibinfo {volume} {12}},\ \bibinfo {pages} {139} (\bibinfo {year} {2016})}\BibitemShut {NoStop}%
\bibitem [{\citenamefont {de~la Barrera}\ \emph {et~al.}(2018)\citenamefont {de~la Barrera}, \citenamefont {Sinko}, \citenamefont {Gopalan}, \citenamefont {Sivadas}, \citenamefont {Seyler}, \citenamefont {Watanabe}, \citenamefont {Taniguchi}, \citenamefont {Tsen}, \citenamefont {Xu}, \citenamefont {Xiao},\ and\ \citenamefont {Hunt}}]{delaBarrera2018}%
  \BibitemOpen
  \bibfield  {author} {\bibinfo {author} {\bibfnamefont {S.~C.}\ \bibnamefont {de~la Barrera}}, \bibinfo {author} {\bibfnamefont {M.~R.}\ \bibnamefont {Sinko}}, \bibinfo {author} {\bibfnamefont {D.~P.}\ \bibnamefont {Gopalan}}, \bibinfo {author} {\bibfnamefont {N.}~\bibnamefont {Sivadas}}, \bibinfo {author} {\bibfnamefont {K.~L.}\ \bibnamefont {Seyler}}, \bibinfo {author} {\bibfnamefont {K.}~\bibnamefont {Watanabe}}, \bibinfo {author} {\bibfnamefont {T.}~\bibnamefont {Taniguchi}}, \bibinfo {author} {\bibfnamefont {A.~W.}\ \bibnamefont {Tsen}}, \bibinfo {author} {\bibfnamefont {X.}~\bibnamefont {Xu}}, \bibinfo {author} {\bibfnamefont {D.}~\bibnamefont {Xiao}},\ and\ \bibinfo {author} {\bibfnamefont {B.~M.}\ \bibnamefont {Hunt}},\ }\bibfield  {title} {\bibinfo {title} {Tuning ising superconductivity with layer and spin--orbit coupling in two-dimensional transition-metal dichalcogenides},\ }\href {https://doi.org/10.1038/s41467-018-03888-4} {\bibfield  {journal} {\bibinfo  {journal} {Nature Communications}\
  }\textbf {\bibinfo {volume} {9}},\ \bibinfo {pages} {1427} (\bibinfo {year} {2018})}\BibitemShut {NoStop}%
\bibitem [{\citenamefont {Saito}\ \emph {et~al.}(2016)\citenamefont {Saito}, \citenamefont {Nakamura}, \citenamefont {Bahramy}, \citenamefont {Kohama}, \citenamefont {Ye}, \citenamefont {Kasahara}, \citenamefont {Nakagawa}, \citenamefont {Onga}, \citenamefont {Tokunaga}, \citenamefont {Nojima}, \citenamefont {Yanase},\ and\ \citenamefont {Iwasa}}]{Saito2016}%
  \BibitemOpen
  \bibfield  {author} {\bibinfo {author} {\bibfnamefont {Y.}~\bibnamefont {Saito}}, \bibinfo {author} {\bibfnamefont {Y.}~\bibnamefont {Nakamura}}, \bibinfo {author} {\bibfnamefont {M.~S.}\ \bibnamefont {Bahramy}}, \bibinfo {author} {\bibfnamefont {Y.}~\bibnamefont {Kohama}}, \bibinfo {author} {\bibfnamefont {J.}~\bibnamefont {Ye}}, \bibinfo {author} {\bibfnamefont {Y.}~\bibnamefont {Kasahara}}, \bibinfo {author} {\bibfnamefont {Y.}~\bibnamefont {Nakagawa}}, \bibinfo {author} {\bibfnamefont {M.}~\bibnamefont {Onga}}, \bibinfo {author} {\bibfnamefont {M.}~\bibnamefont {Tokunaga}}, \bibinfo {author} {\bibfnamefont {T.}~\bibnamefont {Nojima}}, \bibinfo {author} {\bibfnamefont {Y.}~\bibnamefont {Yanase}},\ and\ \bibinfo {author} {\bibfnamefont {Y.}~\bibnamefont {Iwasa}},\ }\bibfield  {title} {\bibinfo {title} {Superconductivity protected by spin--valley locking in ion-gated mos2},\ }\href {https://doi.org/10.1038/nphys3580} {\bibfield  {journal} {\bibinfo  {journal} {Nature Physics}\ }\textbf {\bibinfo {volume}
  {12}},\ \bibinfo {pages} {144} (\bibinfo {year} {2016})}\BibitemShut {NoStop}%
\bibitem [{\citenamefont {Dvir}\ \emph {et~al.}(2018)\citenamefont {Dvir}, \citenamefont {Massee}, \citenamefont {Attias}, \citenamefont {Khodas}, \citenamefont {Aprili}, \citenamefont {Quay},\ and\ \citenamefont {Steinberg}}]{Dvir2018}%
  \BibitemOpen
  \bibfield  {author} {\bibinfo {author} {\bibfnamefont {T.}~\bibnamefont {Dvir}}, \bibinfo {author} {\bibfnamefont {F.}~\bibnamefont {Massee}}, \bibinfo {author} {\bibfnamefont {L.}~\bibnamefont {Attias}}, \bibinfo {author} {\bibfnamefont {M.}~\bibnamefont {Khodas}}, \bibinfo {author} {\bibfnamefont {M.}~\bibnamefont {Aprili}}, \bibinfo {author} {\bibfnamefont {C.~H.~L.}\ \bibnamefont {Quay}},\ and\ \bibinfo {author} {\bibfnamefont {H.}~\bibnamefont {Steinberg}},\ }\bibfield  {title} {\bibinfo {title} {Spectroscopy of bulk and few-layer superconducting nbse2 with van der waals tunnel junctions},\ }\href {https://doi.org/10.1038/s41467-018-03000-w} {\bibfield  {journal} {\bibinfo  {journal} {Nature Communications}\ }\textbf {\bibinfo {volume} {9}},\ \bibinfo {pages} {598} (\bibinfo {year} {2018})}\BibitemShut {NoStop}%
\bibitem [{\citenamefont {Sohn}\ \emph {et~al.}(2018)\citenamefont {Sohn}, \citenamefont {Xi}, \citenamefont {He}, \citenamefont {Jiang}, \citenamefont {Wang}, \citenamefont {Kang}, \citenamefont {Park}, \citenamefont {Berger}, \citenamefont {Forr{\'o}}, \citenamefont {Law}, \citenamefont {Shan},\ and\ \citenamefont {Mak}}]{Sohn2018}%
  \BibitemOpen
  \bibfield  {author} {\bibinfo {author} {\bibfnamefont {E.}~\bibnamefont {Sohn}}, \bibinfo {author} {\bibfnamefont {X.}~\bibnamefont {Xi}}, \bibinfo {author} {\bibfnamefont {W.-Y.}\ \bibnamefont {He}}, \bibinfo {author} {\bibfnamefont {S.}~\bibnamefont {Jiang}}, \bibinfo {author} {\bibfnamefont {Z.}~\bibnamefont {Wang}}, \bibinfo {author} {\bibfnamefont {K.}~\bibnamefont {Kang}}, \bibinfo {author} {\bibfnamefont {J.-H.}\ \bibnamefont {Park}}, \bibinfo {author} {\bibfnamefont {H.}~\bibnamefont {Berger}}, \bibinfo {author} {\bibfnamefont {L.}~\bibnamefont {Forr{\'o}}}, \bibinfo {author} {\bibfnamefont {K.~T.}\ \bibnamefont {Law}}, \bibinfo {author} {\bibfnamefont {J.}~\bibnamefont {Shan}},\ and\ \bibinfo {author} {\bibfnamefont {K.~F.}\ \bibnamefont {Mak}},\ }\bibfield  {title} {\bibinfo {title} {An unusual continuous paramagnetic-limited superconducting phase transition in 2d nbse2},\ }\href {https://doi.org/10.1038/s41563-018-0061-1} {\bibfield  {journal} {\bibinfo  {journal} {Nature Materials}\ }\textbf
  {\bibinfo {volume} {17}},\ \bibinfo {pages} {504} (\bibinfo {year} {2018})}\BibitemShut {NoStop}%
\bibitem [{\citenamefont {Ma}\ \emph {et~al.}(2012)\citenamefont {Ma}, \citenamefont {Dai}, \citenamefont {Guo}, \citenamefont {Niu}, \citenamefont {Zhu},\ and\ \citenamefont {Huang}}]{Ma2012}%
  \BibitemOpen
  \bibfield  {author} {\bibinfo {author} {\bibfnamefont {Y.}~\bibnamefont {Ma}}, \bibinfo {author} {\bibfnamefont {Y.}~\bibnamefont {Dai}}, \bibinfo {author} {\bibfnamefont {M.}~\bibnamefont {Guo}}, \bibinfo {author} {\bibfnamefont {C.}~\bibnamefont {Niu}}, \bibinfo {author} {\bibfnamefont {Y.}~\bibnamefont {Zhu}},\ and\ \bibinfo {author} {\bibfnamefont {B.}~\bibnamefont {Huang}},\ }\bibfield  {title} {\bibinfo {title} {Evidence of the existence of magnetism in pristine vx2 monolayers (x = s, se) and their strain-induced tunable magnetic properties},\ }\href {https://doi.org/10.1021/nn204667z} {\bibfield  {journal} {\bibinfo  {journal} {ACS Nano}\ }\textbf {\bibinfo {volume} {6}},\ \bibinfo {pages} {1695} (\bibinfo {year} {2012})}\BibitemShut {NoStop}%
\bibitem [{\citenamefont {Bonilla}\ \emph {et~al.}(2018)\citenamefont {Bonilla}, \citenamefont {Kolekar}, \citenamefont {Ma}, \citenamefont {Diaz}, \citenamefont {Kalappattil}, \citenamefont {Das}, \citenamefont {Eggers}, \citenamefont {Gutierrez}, \citenamefont {Phan},\ and\ \citenamefont {Batzill}}]{Bonilla2018}%
  \BibitemOpen
  \bibfield  {author} {\bibinfo {author} {\bibfnamefont {M.}~\bibnamefont {Bonilla}}, \bibinfo {author} {\bibfnamefont {S.}~\bibnamefont {Kolekar}}, \bibinfo {author} {\bibfnamefont {Y.}~\bibnamefont {Ma}}, \bibinfo {author} {\bibfnamefont {H.~C.}\ \bibnamefont {Diaz}}, \bibinfo {author} {\bibfnamefont {V.}~\bibnamefont {Kalappattil}}, \bibinfo {author} {\bibfnamefont {R.}~\bibnamefont {Das}}, \bibinfo {author} {\bibfnamefont {T.}~\bibnamefont {Eggers}}, \bibinfo {author} {\bibfnamefont {H.~R.}\ \bibnamefont {Gutierrez}}, \bibinfo {author} {\bibfnamefont {M.-H.}\ \bibnamefont {Phan}},\ and\ \bibinfo {author} {\bibfnamefont {M.}~\bibnamefont {Batzill}},\ }\bibfield  {title} {\bibinfo {title} {Strong room-temperature ferromagnetism in vse2 monolayers on van der waals substrates},\ }\href {https://doi.org/10.1038/s41565-018-0063-9} {\bibfield  {journal} {\bibinfo  {journal} {Nature Nanotechnology}\ }\textbf {\bibinfo {volume} {13}},\ \bibinfo {pages} {289} (\bibinfo {year} {2018})}\BibitemShut {NoStop}%
\bibitem [{\citenamefont {Yu}\ \emph {et~al.}(2019)\citenamefont {Yu}, \citenamefont {Li}, \citenamefont {Herng}, \citenamefont {Wang}, \citenamefont {Zhao}, \citenamefont {Chi}, \citenamefont {Fu}, \citenamefont {Abdelwahab}, \citenamefont {Zhou}, \citenamefont {Dan}, \citenamefont {Chen}, \citenamefont {Chen}, \citenamefont {Li}, \citenamefont {Lu}, \citenamefont {Pennycook}, \citenamefont {Feng}, \citenamefont {Ding},\ and\ \citenamefont {Loh}}]{Yu2019}%
  \BibitemOpen
  \bibfield  {author} {\bibinfo {author} {\bibfnamefont {W.}~\bibnamefont {Yu}}, \bibinfo {author} {\bibfnamefont {J.}~\bibnamefont {Li}}, \bibinfo {author} {\bibfnamefont {T.~S.}\ \bibnamefont {Herng}}, \bibinfo {author} {\bibfnamefont {Z.}~\bibnamefont {Wang}}, \bibinfo {author} {\bibfnamefont {X.}~\bibnamefont {Zhao}}, \bibinfo {author} {\bibfnamefont {X.}~\bibnamefont {Chi}}, \bibinfo {author} {\bibfnamefont {W.}~\bibnamefont {Fu}}, \bibinfo {author} {\bibfnamefont {I.}~\bibnamefont {Abdelwahab}}, \bibinfo {author} {\bibfnamefont {J.}~\bibnamefont {Zhou}}, \bibinfo {author} {\bibfnamefont {J.}~\bibnamefont {Dan}}, \bibinfo {author} {\bibfnamefont {Z.}~\bibnamefont {Chen}}, \bibinfo {author} {\bibfnamefont {Z.}~\bibnamefont {Chen}}, \bibinfo {author} {\bibfnamefont {Z.}~\bibnamefont {Li}}, \bibinfo {author} {\bibfnamefont {J.}~\bibnamefont {Lu}}, \bibinfo {author} {\bibfnamefont {S.~J.}\ \bibnamefont {Pennycook}}, \bibinfo {author} {\bibfnamefont {Y.~P.}\ \bibnamefont {Feng}}, \bibinfo {author}
  {\bibfnamefont {J.}~\bibnamefont {Ding}},\ and\ \bibinfo {author} {\bibfnamefont {K.~P.}\ \bibnamefont {Loh}},\ }\bibfield  {title} {\bibinfo {title} {Chemically exfoliated vse2 monolayers with room-temperature ferromagnetism},\ }\href {https://doi.org/10.1002/adma.201903779} {\bibfield  {journal} {\bibinfo  {journal} {Advanced Materials}\ }\textbf {\bibinfo {volume} {31}},\ \bibinfo {pages} {1903779} (\bibinfo {year} {2019})}\BibitemShut {NoStop}%
\bibitem [{\citenamefont {Chua}\ \emph {et~al.}(2020)\citenamefont {Chua}, \citenamefont {Yang}, \citenamefont {He}, \citenamefont {Yu}, \citenamefont {Yu}, \citenamefont {Bussolotti}, \citenamefont {Wong}, \citenamefont {Loh}, \citenamefont {Breese}, \citenamefont {Goh}, \citenamefont {Huang},\ and\ \citenamefont {Wee}}]{Chua2020}%
  \BibitemOpen
  \bibfield  {author} {\bibinfo {author} {\bibfnamefont {R.}~\bibnamefont {Chua}}, \bibinfo {author} {\bibfnamefont {J.}~\bibnamefont {Yang}}, \bibinfo {author} {\bibfnamefont {X.}~\bibnamefont {He}}, \bibinfo {author} {\bibfnamefont {X.}~\bibnamefont {Yu}}, \bibinfo {author} {\bibfnamefont {W.}~\bibnamefont {Yu}}, \bibinfo {author} {\bibfnamefont {F.}~\bibnamefont {Bussolotti}}, \bibinfo {author} {\bibfnamefont {P.~K.~J.}\ \bibnamefont {Wong}}, \bibinfo {author} {\bibfnamefont {K.~P.}\ \bibnamefont {Loh}}, \bibinfo {author} {\bibfnamefont {M.~B.~H.}\ \bibnamefont {Breese}}, \bibinfo {author} {\bibfnamefont {K.~E.~J.}\ \bibnamefont {Goh}}, \bibinfo {author} {\bibfnamefont {Y.~L.}\ \bibnamefont {Huang}},\ and\ \bibinfo {author} {\bibfnamefont {A.~T.~S.}\ \bibnamefont {Wee}},\ }\bibfield  {title} {\bibinfo {title} {Can reconstructed se-deficient line defects in monolayer vse2 induce magnetism?},\ }\href {https://doi.org/10.1002/adma.202000693} {\bibfield  {journal} {\bibinfo  {journal} {Advanced Materials}\
  }\textbf {\bibinfo {volume} {32}},\ \bibinfo {pages} {2000693} (\bibinfo {year} {2020})}\BibitemShut {NoStop}%
\bibitem [{\citenamefont {Huang}\ \emph {et~al.}(2023)\citenamefont {Huang}, \citenamefont {Jiang}, \citenamefont {Huang},\ and\ \citenamefont {Li}}]{Huang2023}%
  \BibitemOpen
  \bibfield  {author} {\bibinfo {author} {\bibfnamefont {X.}~\bibnamefont {Huang}}, \bibinfo {author} {\bibfnamefont {X.}~\bibnamefont {Jiang}}, \bibinfo {author} {\bibfnamefont {B.}~\bibnamefont {Huang}},\ and\ \bibinfo {author} {\bibfnamefont {Z.}~\bibnamefont {Li}},\ }\bibfield  {title} {\bibinfo {title} {Nonlocal optical conductivity of fermi surface nesting materials},\ }\href {https://doi.org/10.1007/s11433-022-2035-7} {\bibfield  {journal} {\bibinfo  {journal} {Science China Physics, Mechanics {\&} Astronomy}\ }\textbf {\bibinfo {volume} {66}},\ \bibinfo {pages} {247011} (\bibinfo {year} {2023})}\BibitemShut {NoStop}%
\end{thebibliography}%

\end{document}